# On the Modified Einstein-Laub and Modified Chu Optical Force Formulations

M.R.C. Mahdy[1,2*], Tianhang Zhang[2], Golam Dastegir Al-Quaderi[3], Hamim Mahmud Rivy[1], Amin Kianinejad[2]

[1]Department of Electrical & Computer Engineering, North South University, Bashundhara, Dhaka 1229, Bangladesh

[2] Department of Electrical and Computer Engineering, National University of Singapore, 4 Engineering Drive 3, Singapore 117583

[3] Department of Physics, University of Dhaka, Dhaka-1000, Bangladesh

[*] Corresponding author: mahdy.chowdhury@northsouth.edu

In several experiments involving material background, it has been observed that the Chu, Einstein-Laub and Ampere formulations of optical force lead to either different optical forces or wrong total optical force. In order to identify the exact reason behind such significant disagreements, we investigate the optical force in a number of tractor beam and lateral force experiments. We demonstrate that the modified Einstein-Laub or modified Chu formulations, obtained from two 'mathematical consistency conditions' of force calculation, give the time-averaged force that agrees with the experiments. We consider both the chiral and achiral objects embedded in complex material backgrounds. Though the distinct formulations of optical force have been made 'mathematically equivalent' in this work; the aspect of physical consistency of these distinct optical force formulations have also been investigated. It is known that Minkowski's theory suggests zero bulk force inside a lossless object for which we still do not have any experimental verification. In contrast, both modified Einstein-Laub and modified Chu force formulations suggest non-zero bulk force inside a lossless object. Hence, for a future resolution of this discrepancy, we also suggest a possible experiment to investigate the bulk force and to check the validity of these distinct formulations.

## I. INTRODUCTION

Recent experiments of optical manipulation, such as tractor beam and lateral force-based investigations [1-15], have opened up a novel path of study promising a possible resolution of the ongoing debates [16-19] on time-averaged total forces. Since the tractor beam and lateral force experiments involve the material background [1, 2, 9, 11, 12], these will help to understand the persistently debated roles of different stress tensors (STs), optical forces and photon momenta, i.e. the Abraham-Minkowski controversy [16-19]. The employment of an inappropriate approach may lead to a pushing force or inconsistent lateral force, instead of the experimentally observed pulling force (or consistent lateral force). This article focuses mainly these aspects of the tractor beam and lateral force experiments.

It is commonly believed and expected that all the volume force formulations [18, 20] [i.e. Minkowski, Abraham, Chu, Einstein-Laub (EL) and Ampere/Nelson as shown in Table 1] should lead to the same time averaged total optical force. But this idea has been invalidated in some recent works [3, 19] considering several experimental set-ups [19] involving a material background. For example, the well-known Chu and Einstein-Laub optical force formulations [18-20] lead to incorrect (either in magnitude or direction according to [19] and [21]) time-averaged total forces for the Hakim-Higham experiment [21,22], the Jones-Richard experiment [23], Jones-Leslie experiment [24], 1973 Ashkin-Dziedzic experiment [25], and those on an half-immersed object, and so on.

When a material background is involved, several complexities arise to yield the time averaged correct total force. These are:

(i) In order to calculate the total optical force on an embedded object, two alternative methods, (i.e., '*gap method*' and '*no gap method*'), are applied [20, 26-30]. (a) In the '*gap method*' [20, 28-30], a very small gap is introduced between the scatterer and the background to yield the force as shown in Fig. 1(a). This method was introduced to make all distinct formulations [cf. Table 1] equivalent to each other even if the background medium is *not* air or vacuum. (b) In the '*no gap method*' [1-3,12,19,26,27] optical force is calculated without introducing any gap between the embedded scatterer and the material background as shown in Fig. 1(b) and (c). In this article we investigate whether all the distinct force formulations give the same result in both the aforementioned methods and if not what is the resolution to possible inconsistencies?

(ii) For the '*no gap*' method, although Minkowski's and Abraham's force formulations lead to a fully consistent time-averaged total force (correct both in magnitude and sign) Chu and Einstein-Laub forces lead to wrong results for several real experiments [19, 21]. Hence, for the 'no gap method', is it possible to make distinct formulations 'mathematically equivalent' to each other for both chiral [31-37] and achiral [2, 3, 19, 21, 39-44] objects embedded in a medium?

This work attempts to solve the aforementioned complexities [3, 19, 21, 38] by introducing the Modified Einstein-Laub (MEL) and Modified Chu (MChu) force formulations. Throughout this paper we refer to 'exterior' or 'outside' magnitudes those evaluated outside the volume of the embedded macroscopic object, while by 'interior' or 'inside' we refer to those quantities which are evaluated inside the object [cf. the force calculation process in Fig.1 (a) - (c)].

Two notable reasons for which this work is important are:

**(1)** For an object, the total force is usually described as [28-30,38]: $\langle \boldsymbol{F}^{Total} \rangle = \langle \boldsymbol{F}^{Bulk} \rangle (\text{in}) + \langle \boldsymbol{F}^{Surface} \rangle$. where $\langle \boldsymbol{F}^{Bulk} \rangle (\text{in})$ is the time averaged force inside an object, and $\langle \boldsymbol{F}^{Surface} \rangle$ is the surface force. For a lossless object, Minkowski's theory suggests that: $\langle \boldsymbol{F}^{Bulk} \rangle (\text{in}) = 0$ and hence [cf. Eq (58) given in ref. [45]]: $\langle \boldsymbol{F}^{Total} \rangle = \langle \boldsymbol{F}^{Surface} \rangle$. In contrast, for the proposed MEL or MChu [26] formulations $\langle \boldsymbol{F}^{Bulk} \rangle (\text{in}) \neq 0$ [26, 27] and hence: $\langle \boldsymbol{F}^{Total} \rangle = \langle \boldsymbol{F}^{Bulk} \rangle (\text{in}) + \langle \boldsymbol{F}^{Surface} \rangle$.

It will be discussed in this article: $\langle \boldsymbol{F}^{Total} \rangle$ gives the exact same time averaged force for these three distinct methods: Minkowski force, MEL and MChu force; though the volumetric force distribution process is fully different for them (as discussed: Minkowski force's total force arises fully from the surface of a lossless object).

So far no such experiment has been conducted, which can verify the presence of bulk force inside a material medium due to technological constraints. If in the future, any experiment detects the existence of bulk force, it would require reconsidering the accepted Minkowski theory and demand examining the MEL or MChu formulations.

**(2)** The volumetric force of MChu formulation is widely applied [39] for optical force calculations:

$$\boldsymbol{F}_v = \varepsilon_b (\nabla \cdot \boldsymbol{E}) \boldsymbol{E} + i\omega (\varepsilon_s - \varepsilon_b) \boldsymbol{E}_{in} \times \boldsymbol{B}.$$

Where $\varepsilon_b$ is the background permittivity (where the object is embedded), but $\varepsilon_s$ is the permittivity of the embedded object. It means that: though the force is calculated for an embedded object, some mathematical parameters (i.e. background permittivity) are used in the mathematical formula of optical force calculation of MChu. For example- if a Silica object is fully immersed in water; $\varepsilon_b$ is the permittivity of water but $\varepsilon_s$ is the permittivity of Silica (integration boundary of the force calculation should be given enclosing only the Silica object). Using the widely applied full-wave simulation technique presented in [39], several works [40, 42] have been recently reported in literature predicting accurate results of time-averaged total optical force, which (the MChu formulation [26, 27, 40, 42]) leads to the exactly same time averaged total force of Minkowski force formulation. In this article, we have shown that MEL formulations are also exactly mathematically equivalent to the Mchu and Minkowski formulations.

However, the detail theoretical (underlying physics and other issues) investigation in favor of both MEL and MChu formulation is still not available in literature. For example- the associated stress tensor (time average scenario) formulations related to MChu and MEL formulations have not been discussed in detail in literature. This work attempts to present such a detail theoretical investigation (both analytically and based on full wave simulations) along with the possible drawbacks (the 'underlying physics' of such formulations) of these two formulations.

Notably, based on analytical calculations, full wave simulation based results and the experiments performed so far, in this article we have tried to figure out the fact whether MEL and MChu formulations can really be widely used as an alternative to Minkowski force law. After a detail investigation, we have concluded that MEL and MChu formulations can certainly be used as very efficient mathematical toolkits to yield the time averaged total force on an embedded object. In addition, we have also investigated the physical aspects of these efficient mathematical toolkits: the MEL and MChu formulations.

**Table 1: Well-Known optical force formulations and stress tensors (STs) for lossless objects**

| Name of Force | External ST** | Internal ST** | Bulk force density (not time averaged) | Surface force density | Bulk force $\langle F_{Bulk} \rangle$ | Surf. force $\langle F_{Surf} \rangle$ |
|---|---|---|---|---|---|---|
| Minkowski | $\bar{\bar{T}} = D_{out}E_{out} + B_{out}H_{out} - \frac{1}{2}(B_{out} \times H_{out} + D_{out} \times E_{out})\bar{\bar{I}}$ | $\bar{\bar{T}} = D_{in}E_{in} + B_{in}H_{in} - \frac{1}{2}(B_{in} \cdot H_{in} + D_{in} \cdot E_{in})\bar{\bar{I}}$ | 0 (considering homogeneous object) | $f^{Surface} = -[\frac{1}{2}E_{out} \cdot E_{in}\Delta\varepsilon + \frac{1}{2}H_{out} \cdot H_{in}\Delta\mu]_{at\ r=a}$ | 0 | Non-zero(= total time averaged force) |
| Abraham | $\bar{\bar{T}} = \frac{1}{2}\begin{bmatrix} D_{out}E_{out} + E_{out}D_{out} \\ + B_{out}H_{out} \\ + H_{out}B_{out} \\ -\begin{pmatrix} B_{out} \cdot H_{out} \\ +D_{out} \cdot E_{out} \end{pmatrix}\bar{\bar{I}} \end{bmatrix}$ | $\bar{\bar{T}} = \frac{1}{2}\begin{bmatrix} D_{in}E_{in} + E_{in}D_{in} \\ +B_{in}H_{in} + H_{in}B_{in} \\ -\begin{pmatrix} B_{in} \cdot H_{in} \\ +D_{in} \cdot E_{in} \end{pmatrix}\bar{\bar{I}} \end{bmatrix}$ | $f^{Bulk}(in) = (n^2-1) \times \frac{\partial}{\partial t}\left(\frac{E_{in} \times H_{in}}{c^2}\right)$ | $f^{Surface} = -[\frac{1}{2}E_{out} \cdot E_{in}\Delta\varepsilon + \frac{1}{2}H_{out} \cdot H_{in}\Delta\mu]_{at\ r=a}$ | 0 | Non-zero(= total time averaged force) |
| Einstein-Laub | $\bar{\bar{T}} = D_{out}E_{out} + B_{out}H_{out} - \frac{1}{2}(\mu_o H_{out} \cdot H_{out} + \varepsilon_0 E_{out} \cdot E_{out})\bar{\bar{I}}$ | $\bar{\bar{T}} = D_{in}E_{in} + B_{in}H_{in} - \frac{1}{2}(\mu_o H_{in} \cdot H_{in} + \varepsilon_0 E_{in} \cdot E_{in})\bar{\bar{I}}$ | $f^{Bulk}(in) = (P_{in} \cdot \nabla)E_{in} + (M_{in} \cdot \nabla)H_{in} + \frac{\partial P_{in}}{\partial t} \times \mu_o H_{in} - \frac{\partial M_{in}}{\partial t} \times \varepsilon_0 E_{in}$ | $f^{Surface} = \{\varepsilon_0(E_{out} - E_{in}) \cdot \hat{n}\}\left(\frac{E_{out} - E_{in}}{2}\right)_{at\ r=a} + \{\mu_0(H_{out} - H_{in}) \cdot \hat{n}\}\left(\frac{H_{out} - H_{in}}{2}\right)_{at\ r=a}$; [28],[30],[38] | Non-zero | Non-zero |
| Chu | $\bar{\bar{T}} = \varepsilon_0 E_{out}E_{out} + \mu_0 H_{out}H_{out} - \frac{1}{2}(\mu_o H_{out} \cdot H_{out} + \varepsilon_0 E_{out} \cdot E_{out})\bar{\bar{I}}$ | $\bar{\bar{T}} = \varepsilon_0 E_{in}E_{in} + \mu_0 H_{in}H_{in} - \frac{1}{2}(\mu_o H_{in} \cdot H_{in} + \varepsilon_0 E_{in} \cdot E_{in})\bar{\bar{I}}$ | $f^{Bulk}(in) = -(\nabla \cdot P_{in})E_{in} - (\nabla \cdot M_{in})H_{in} + \frac{\partial P_{in}}{\partial t} \times \mu_0 H_{in} - \frac{\partial M_{in}}{\partial t} \times \varepsilon_0 E_{in}$ | $f^{Surface} = [\{\varepsilon_0(E_{out} - E_{in}) \cdot \hat{n}\}\left(\frac{E_{out} + E_{in}}{2}\right)_{at\ r=a} + \{\mu_0(H_{out} - H_{in}) \cdot \hat{n}\}\left(\frac{H_{out} + H_{in}}{2}\right)_{at\ r=a}]$; [28],[30],[38] | Non-zero | Non-zero |
| Ampere (for dielectric objects only) | $\bar{\bar{T}} = \varepsilon_0 E_{out}E_{out} + \mu_0^{-1}B_{out}B_{out} - \frac{1}{2}\left(\mu_0^{-1}B_{out}^2 + \varepsilon_0 E_{out}^2\right)\bar{\bar{I}}$ | $\bar{\bar{T}} = \varepsilon_0 E_{in}E_{in} + \mu_0^{-1}B_{in}B_{in} - \frac{1}{2}\left(\mu_0^{-1}B_{in}^2 + \varepsilon_0 E_{in}^2\right)\bar{\bar{I}}$ | $f^{Bulk}(in) = -(\nabla \cdot P_{in})E_{in} + \frac{\partial P_{in}}{\partial t} \times B_{in} + (\nabla \times M_{in}) \times B_{in}$ | $f^{Surface} = [\{\varepsilon_0(E_{out} - E_{in}) \cdot \hat{n}\}\left(\frac{E_{out} + E_{in}}{2}\right)_{at\ r=a}]$ (for dielectric objects only) | Non-zero | Non-zero |

* 'out' stands for the magnitudes those evaluated outside the volume of a macroscopic object, while 'in' stands for those quantities inside the object volume [cf. the force calculation process in Fig.1]. *P* and *M* are induced polarization and magnetization of the scatterer respectively.

## II. THEORY: TWO 'MATHEMATICAL CONSISTENCY CONDITIONS'

One of the consistency conditions of any optical force formulation is the equation of continuity of force density given by: [18, 19, 21, 45]:

$$\nabla \cdot \overline{\overline{T}} = f + \frac{\partial}{\partial t} G$$

Which leads to a time averaged force on an object:

$$\langle F \rangle = \oiint \langle \overline{\overline{T}} \rangle \cdot ds = \int \langle f \rangle dv.$$

Two more physically plausible and obvious conditions (which we call the consistency conditions 'CI' and 'CII') must be satisfied by any correct time averaged total force where the external and internal electromagnetic fields of an object with radius $r = a$ are used:

1. Consistency condition- 1 ('CI'):

$$[\overline{\overline{T}}(\text{out}) - \overline{\overline{T}}(\text{in})] \cdot \hat{n} = f_1^{\text{Surface}} = f_2^{\text{Surface}} = f^{\text{Surface}}$$

- Here, $\overline{\overline{T}}(\text{out})$ is an external stress tensor, applied at $r = a^+$ using exterior fields of the object.

   $\overline{\overline{T}}(\text{in})$ is the internal stress tensor (inside the object), applied at $r = a^-$, [cf. Fig. 1 (b) and (c)], employing only the interior field of the scattering particle and $f$ (in) is its corresponding volume force density with $\langle F^{\text{Bulk}} \rangle (\text{in}) \neq 0$.

Thus, on the boundary $r = a$ of any object, one should determine the quantity: $[\overline{\overline{T}}(\text{out}) - \overline{\overline{T}}(\text{in})] \cdot \hat{n} = f_1^{\text{Surface}}$ where $\hat{n}$ is the local unit outward normal of the object surface and then one should independently calculate the surface force density, $f_2^{\text{Surface}}$ from the volume force density $f$ (in) by applying the appropriate boundary conditions at $r = a$. These independently calculated two quantities should be same [both in magnitude and direction]: $f_1^{\text{Surface}} = f_2^{\text{Surface}} = f^{\text{Surface}}$,"

2. Consistency condition- 2 ('CII'):

$$\oiint \langle \overline{\overline{T}}(out) \rangle \cdot ds = \langle F^{Total}(Consistent) \rangle = \langle F^{Bulk} \rangle (in) + \langle F^{Surface} \rangle$$

- In other words, the independently calculated external time averaged force $\oint \langle \bar{\bar{T}}(out) \rangle \cdot ds$ should be exactly equal to the independently calculated internal bulk force plus the surface force of an object.

In [38], the process of obtaining $f_2^{Surface}$ in 'CI', has been shown, using a volumetric bulk force density, $f$ (in) - stemming from the Chu and Einstein-Laub force, and by considering only air as the background. In our work, we attempt the same approach for a volumetric force ($f$), where an additional contribution comes from the stress tensors ($[\bar{\bar{T}}(\text{out}) - \bar{\bar{T}}(\text{in})] \cdot \hat{n} = f_1^{Surface}$) due to the the presence of a material background.

- We have shown that (cf supplement S1) when the background is air, the difference of the external ST (notice that then all STs are the same) and the internal ST of Chu (and of Einstein-Laub) at the object boundary is in agreement with the independently calculated surface force given by the volumetric formulation of Chu (and Einstein-Laub). In fact, this conclusion is true for all the well-known force density laws and their corresponding internals STs, as discussed in Table 2, when the background is air/vacuum instead of any other denser material. *In contrast, it is important to note that: when the background is a material medium, $\bar{\bar{T}}(\text{out})$ and $\bar{\bar{T}}(\text{in})$ cannot be arbitrary STs which satisfy 'CI' and 'CII' simultaneously.* Through analytical calculations, we have pointed out in Table 2 that some of the candidates in the sets of $\bar{\bar{T}}(\text{out})$, $\bar{\bar{T}}(\text{in})$ and the corresponding $f$ (in) *satisfy 'CI' and 'CII' simultaneously while some do not.* As an example, if an object is embedded in a material medium, 'the difference of $\bar{\bar{T}}(\text{out})$ and $\bar{\bar{T}}(\text{in})$' of the well–known EL stress tensor does not match with the independently calculated surface force obtained from the $f$ (in) of the EL force density. This problem is solved by the forthcoming proposed MEL (and Mchu) formulations [cf. Table A1 and A2 in APPENDIX A].

**Table 2: Differences of 'Consistency Conditions' for air and material backgrounds**

| Name of Force (x) | 'CII' [for air background] $\oint \langle \bar{\bar{T}}_{Vacuum}(out) \rangle \cdot ds$ $\langle F_x^{Total}(Consistent) \rangle$ $= \langle F_x^{Bulk} \rangle (in)$ $+ \langle F_x^{Surface} \rangle$ satisfied? | 'CI' [for air background]** $[\bar{\bar{T}}_{Vacuum}(out)$ $-\bar{\bar{T}}_x(in)] \cdot \hat{n};_{at\ r=a}$ $= f_x^{Surface}$ satisfied? | 'CII' [for material background]**** $\oint \langle \bar{\bar{T}}_{Mink}(out) \rangle \cdot ds =$ $\langle F_x^{Total}(Consistent) \rangle|$ $= \langle F_x^{Bulk} \rangle (in)$ $+ \langle F_x^{Surface} \rangle$ satisfied? | 'CI' [for material background]**** $[\bar{\bar{T}}_{Mink}(out) -$ $\bar{\bar{T}}_x(in)] \cdot \hat{n};_{at\ r=a}$ $= f_x^{Surface}$ satisfied? | Additional condition [for material background] $[\bar{\bar{T}}_x(out) -$ $\bar{\bar{T}}_x(in)] \cdot \hat{n};_{at\ r=a}$ $= f_x^{Surface}$ satisfied? |
|---|---|---|---|---|---|
| Minkowski | Yes | Yes (independent derivation of Helmholtz force density can be found in refs [46] sec 2.21 and [47].) | Yes | Yes | Yes |
| Abraham | Yes | Yes | Yes | Yes | Yes |
| Einstein-Laub | Yes | Yes | No [though sign of total force is consistent [19], the magnitudes are always different than the force magnitudes of Mink. and Abraham [19] (especially for the cases of refs [23] and [24] according to [19]). Also see the additional refs [21] (i.e. Hakim –Higham experiment), [48], [49] where EL formulations fail]. | No | No |
| Chu | Yes | Yes | No [19] | No | Yes |
| Ampere [for dielectric object only] | Yes | Yes | No [19] | No | Yes |

Notes: **(1) $\bar{\bar{T}}_{Vacuum}(out) = D_{out}E_{out} + B_{out}H_{out} - \frac{1}{2}(B_{out} \cdot H_{out} + D_{out} \cdot E_{out})\bar{\bar{I}}$ and ****(2) $\bar{\bar{T}}(out)$ has been considered $\bar{\bar{T}}_{Mink}(out)$ due to three notable reasons. These 'notable' reasons have been discussed in detail in the first sub-section of section III.

## III. RESULTS AND DISCUSSIONS

### IIIA. The consistent external force with the '*no gap method*'

We have investigated the consistency of the total force calculated for one tractor beam experiment [1] and for one on a lateral force [12], using different major external stress tensors, i.e. those due to Minkowski, Chu, Ampere and Einstein-Laub [cf. Fig. 1(b) and (c)]. All 3D simulations throughout this paper have been conducted using an incident power of $0.57\text{mW}/\mu m^2$.

For the '*no gap method*', the most popular and most accurate choice of the stress tensor is the Minkowski stress tensor employed in the external time-averaged total force calculation. This is because of the following reasons:

(i) Reason 1: Relativistic invariance of Minkowski formulation [50,51] which is absent in the other formulations ( Abraham [50,51] and Einstein-Laub force laws [52]), except in Chu [52] and Ampere forces.

(ii) Reason 2: The consistency condition 'CI' remains valid when both $\bar{\bar{T}}(\text{out})$ and $\bar{\bar{T}}(\text{in})$ are taken to be the Minkowski stress tensor. This is not the case for several other formulations (cf. the sixth column in Table 2).

(iii) Reason 3: The consistency condition 'CII' remains valid for the set-ups of all the experiments conducted so far [19,21] when we calculate (i) the total time averaged outside force using the Minkowski stress tensor (satisfying the left side of the equation of 'CII' quoted above) and (ii) independently calculate the surface force using Helmholtz's force law (satisfying the left side of the equation of 'CII' above, though $\langle F_{Mink}^{Bulk} \rangle(in) = \int \langle f_{Helmholtz}^{Bulk} \rangle(in) \cdot dv = \oiint \langle \bar{\bar{T}}_{Mink}(in) \rangle \cdot ds = 0$ for non-absorbing objects).

Hence, the most consistent external time-averaged total outside-force (calculated using the Minkowski stress tensor) can be written as [1-3,12]:

$$\langle F_{Total} \rangle(out) = \oiint \langle \bar{\bar{T}}_{Mink}^{out} \rangle \cdot ds \tag{1a}$$

$$\langle \bar{\bar{T}}_{Mink}^{out} \rangle = \frac{1}{2}\text{Re}\left[ D_{out}E_{out}^* + B_{out}H_{out}^* - \frac{1}{2}\left( B_{out} \cdot H_{out}^* + D_{out} \cdot E_{out}^* \right)\bar{\bar{I}} \right]. \tag{1b}$$

Where the '*out*' stands for fields outside the object, ( at $r=a^+$, if it is a sphere or cylinder of radius $a$, cf. Fig. 1(b)) and $\bar{\bar{I}}$ is the unit dyadic. The electromagnetic fields occurring in Eq. (1b) are the total electric and magnetic fields i.e. the incident plus scattered field by the body.

Without introducing any small gap, in Supplement S2, Fig. 2s(a) –2s(c), we calculate the total outside force using different external stress tensors for the configurations of (i) the two-beamed tractor beam experiment, reported in [1] and (ii) the lateral force experiment [12] as in Fig. 2(a) – (c) .Due to the big size of the object (4500 nm), as was used in the lateral force experiment [12], (required a special technique in the first reference quoted in [12]), we model that set-up with a comparatively small-sized object (cf. the second reference quoted in [12], which addresses a smaller object). We are considering two configurations, namely, a spherical and an elliptical object to check the consistency of the lateral force. It may be noted that for an interfacial tractor beam experiment [2,3], only the external Minkowski stress tensor gives correct results for objects [2,3] modeled as either spherical or elliptical. Though there is no difference among the signs of the total forces, (calculated using the different external stress tensors) for the experiments reported in [1] and [12], their magnitudes are quite different, as shown in Supplement S2, Figs 2s (b), (c) (for two beamed tractor beam experiment) and Figs 2(b), (c) (for the lateral force experiment).

Note that, for the '*no gap method* ' one has: (A) size-based sorting of embedded particles by a two-beam procedure (cf. supplement of ref. [1]), which is predicted using all external stress tensors; and (B) a change of sign of the force for two different handedness of polarizations observed for the lateral force experiment (cf. [12]), which has also been correctly predicted using all external stress tensors for spherical and elliptical objects.

In Fig. 3(a) given in the second reference quoted in [12], for a 1500 nm-sized object, the lateral force has been observed to have a single direction for the single handedness of circularly polarized light, which is consistent with our result. These aforementioned observations will be important for our next investigation i.e. the consistency test of the '*gap method*' for the tractor beam [1,2] and for the lateral force experiment [12].

### IIIB. Inconsistency of the '*gap method*' and the necessity of a modification

If the background is a material medium, the optical force may be calculated by introducing an extremely small gap between the object and the background medium (cf. Fig. 1(a)) [20, 28-30]. In this section, we shall investigate the results obtained using the 'gap *method*' for scatterer embedded in material media (as in [1], [2], and [12]). If a small gap is introduced (cf. Fig. 1(a)), the external vacuum stress tensor (e.g. the Chu stress tensor [20]) that yields the total outside force, should be written in terms of the gap fields ($E_g$, $H_g$) as [20]:

$$\langle F_{Total} \rangle (out) = \oiint \langle \bar{\bar{T}}^{out}_{Vacuum} \rangle \cdot ds, \tag{2a}$$

$$\langle \bar{\bar{T}}^{out}_{Vacuum} \rangle = \frac{1}{2} \text{Re} \left[ \varepsilon_0 E_g E_g^* + \mu_0 H_g H_g^* - \frac{1}{2} \left( \mu_0 H_g \cdot H_g^* + \varepsilon_0 E_g \cdot E_g^* \right) \bar{\bar{I}} \right]. \tag{2b}$$

In Fig. 3(a) we consider an extremely small gap [cf. Fig. 1(a)] between the half immersed scatterer and the water background. By employing the stress tensor of Eq. (2b), we calculate the external force. In this case, instead of the experimentally observed pulling force [2, 3], we obtain a pushing force due to such an extremely small gap (cf in Fig. 3(a)). In contrast to this case, both the '*gap method*' and the '*no gap method*' lead to consistent pulling forces for the two-beamed tractor beam experiment [1] (cf. Fig. 3(b)).

For the lateral force experiment reported in [12], for a spherical object, the total force is in good agreement (cf. Fig. 3(c)) with the calculated forces with a '*no gap method*', (cf. Fig. 2 (b)) whereas for an elliptical object the sign differs (cf. Fig. 3(d)) in comparison with the '*no gap method*' [cf. Fig. 2(c)]. Thus, we have a wrong result with the 'gap method' when the background is inhomogeneous [2,3,12], or, in general, the object is not symmetric in shape (cf. Fig. 1(b)). This clearly violates the consistency condition 'CII'. In supplement S1 we discuss in detail why the '*gap method*' cannot constitute a general way to obtain the total force. For the so far reported experiments (especially those experiments with inhomogeneous (or heterogeneous) backgrounds) due to the violation of 'CI', the 'gap method' fails. In fact, the linear increase of photon momentum exactly at the interface of two different medium (due to reduced impedance mismatch [53]) cannot be accounted properly, if one introduces such an artificial gap. This ultimately leads to the wrong total force for the 'gap method'.

However, the failure of Einstein-Laub and Chu formulations with both the '*gap method*' and '*no gap method*' (discussed previously) can be overcome by the modification of the well-known Chu or Einstein-Laub formulations, which (the modification) will be presented in the next section. It should be noted that the forthcoming modified EL and Chu formulations are not applicable for 'gap method'. Both of them are strictly applicable only for 'no gap method' [similar to the Minkowski's formulations] to yield the correct time-averaged force. So, we are ultimately going to deal the question: whether MEL (or MChu) formulation of 'no gap method' is a correct alternative of Minkowski formulation of 'no gap method' or not. This issue will also be discussed in the later part of this article.

## IV. MODIFICATION OF THE EINSTEIN-LAUB AND CHU FORMULATIONS

(1) The very first difference between the well-accepted Minkowski's force formulations and EL force formulations arises from the bulk part of the force. It is discussed already in the introduction of this article: in the Minkowski formulation, the bulk force is zero while that using Einstein-Laub formulation is non-zero. In this section, we put forward a modified version of the Einstein-Laub formulation, called the MEL formulation for '*no gap method*'. As discussed in APPENDIX B, the achiral time-averaged MEL stress tensor (which should yield the bulk force on an embedded object) just inside an embedded object (i.e. at $r=a^-$) should be written as:

$$\left\langle \overline{\overline{T}}_{MEL(j)}^{Bulk} \right\rangle (in) = D_{in} E_{in}^* + B_{in} H_{in}^* - \frac{1}{2}\left( (\frac{\mu_{b\,(j)}}{\mu_s})\mu_s H_{in}^* \cdot H_{in} + (\frac{\varepsilon_{b(j)}}{\varepsilon_s})\varepsilon_s E_{in}^* \cdot E_{in} \right)\overline{\overline{I}}. \tag{3}$$

where $j=1,2,3,....,N$ represents the number of background regions sharing interface with the object, (cf. Figs. 1(b), (c), Fig. 5(a)), $\varepsilon_b$ and $\mu_b$ are the background permittivity and permeability, respectively and $\varepsilon_s$ and $\mu_s$ stand for the permittivity and permeability, respectively of the scatterer.

Just for example- if a Silica object is fully immersed in water; $\varepsilon_b$ is the permittivity of water but $\varepsilon_s$ is the permittivity of Silica (integration boundary of the force calculation should be given enclosing only the Silica object). Especially for the internal time-averaged stress tensor calculation (as shown in Eq (3)), the integration boundary should be applied just inside the Silica object (i.e. at $r=a^-$) to yield 'only' the time-averaged 'bulk force' [not for the 'surface force']. From Eq (3), it can be understood that: background permittivity and permeability enter in the MEL stress tensor equation as ratio based unitless quantitities : $(\frac{\varepsilon_{b(j)}}{\varepsilon_s})$ and $(\frac{\mu_{b(j)}}{\mu_s})$. It should also be noted that Eq (3) is a time averaged version of the stress tensor.

We take the permittivities and the permeabilities as constant, i.e. independent of the frequency or other properties of the waves. As shown in APPENDIX B, the achiral MEL stress tensor should be written in the time-averaged form which is related to the time-averaged MEL volume force density:

$$\left\langle f_{MEL(j)}^{Bulk}\right\rangle(in) = \frac{1}{2}\text{Re}\left[\left(P_{Eff(j)}\cdot\nabla\right)E_{in}^* + \left(M_{Eff(j)}\cdot\nabla\right)H_{in}^* - \left(i\omega P_{Eff(j)}\times B_{in}^*\right) + \left(i\omega M_{Eff(j)}\times D_{in}^*\right)\right]. \quad (4)$$

A similarly modified formulation can also be written for the Chu formulation for achiral embedded objects given in [26, 39, 54]. In Eq. (4), the effective polarization [26,27] and magnetization [26] are defined as: $P_{Eff} = (\varepsilon_S - \varepsilon_b)E_{in}$ and $M_{Eff} = (\mu_S - \mu_b)H_{in}$. The total time-averaged Bulk force on the embedded object should be:

$$\left\langle F_{MEL}^{Bulk}\right\rangle(in) = \sum_j \oiint \left\langle \overline{\overline{T}}_{MEL(j)}^{Bulk}\right\rangle(in)\cdot ds_{(j)} = \sum_j \int \left\langle f_{MEL(j)}^{Bulk}\right\rangle(in)\cdot dv_{(j)}. \quad (5)$$

(2) Now, by applying the appropriate boundary conditions (cf. APPENDIX C), the surface force of the modified Einstein-Laub method can be written in two different ways: (i) that calculated using the volume force method of Eq. (4), (as shown in [38] only for a volumetric force), and (ii) that calculated from the difference between the external Minkowski stress tensor and the internal MEL stress tensors in Eq. (3) just at the boundary. These two different ways lead to the same result, (cf. APPENDIX C) and hence fulfills the consistency condition 'CI':

$$f_{MEL}^{Surface} = [\overline{\overline{T}}_{Mink}(out) - \overline{\overline{T}}_{MEL(j)}(in)]\cdot \hat{n}_{r=a}$$
$$= \left\{\varepsilon_{b(j)}(E_{out} - E_{in})\cdot \hat{n}\right\}\left(\frac{E_{out} - E_{in}}{2}\right)_{at\ r=a} + \left\{\mu_{b(j)}(H_{out} - H_{in})\cdot \hat{n}\right\}\left(\frac{H_{out} - H_{in}}{2}\right)_{at\ r=a} \quad (6)$$

Eq. (6) explains why in [27], the time-averaged total force predicted by Minkowski stress tensor is in exact agreement with the MEL volume force reported in [28]. Hence, the total time-averaged force on an embedded generic object according to MEL formulation should finally be written as (also applicable for the modified Chu formulation [26, 39, 54]):

$$\oiint \langle \overline{\overline{T}}_{Mink}^{out} \rangle \cdot ds = \langle F_{MEL}^{Bulk} \rangle (in) + \langle F_{MEL}^{Surface} \rangle \quad (7)$$

Hence, we have analytically shown how both the consistency conditions CI and CII are satisfied in the MEL formulation. The validity of Eq. (7) for the consistency condition, CII is investigated in the next section, both for achiral and chiral objects based on numerical (full wave) simulations for several complicated configurations. The final conclusions based on our results are shortly listed in tables A1 and A2 of APPENDIX A.

It is important to note that if we consider the background medium of an object as air (or vacuum) instead of some other denser medium, the proposed time-averaged modified internal EL stress tensor, surface force and bulk force formulations (cf. Table A1 in APPENDIX A) all transform into the well-known time-averaged internal EL stress tensor, surface force and bulk force respectively (as it should be).

## V.  DETAIL RESULTS AND DISCUSSIONS FOR ACHIRAL AND CHIRAL OBJECTS

From Fig.4 (a) and Fig. 4(b), we see that the total external time-averaged force obtained using the Minkowski stress tensor and the total internal bulk force from MEL stress tensor are in almost full agreement for the two beamed tractor beam experiment of reference [1]. This is due to a very small value of $[1-(\frac{\varepsilon_{b(j)}}{\varepsilon_s})]^2$. However, the difference between the bulk force calculated using the MEL stress tensor and the total force of the external Minkowski stress tensor is clearly non-negligible for the lateral force experiment when $[1-(\frac{\varepsilon_{b(j)}}{\varepsilon_s})]^2$ is not very small, (for example for a TiO$_2$ object embedded in an air-water interface; see Fig. 4(c)). Also note that the bulk force calculated using the MEL stress tensor (cf. Fig. 4(d)), depends on the size or shape of the object. However, by adding the surface force of the achiral MEL to the bulk force (cf. Eqs (6) and (7)), the magnitude of the optical force matches exactly with that of the external time-averaged total force given by the Minkowski stress tensor. In supplement S3 the bulk force calculation using MEL stress tensor has been shown for a Mie, and more complex object, embedded in either a homogeneous, heterogeneous, or bounded background medium. To the best of our knowledge, force calculations in the presence of heterogeneous media have not been previously discussed in the literature. Introduction of chirality into the object has been a key factor for checking the mathematical consistency of the methods to calculate the optical force [31-37]. The detailed mathematical calculations using the Chiral MEL formulations are presented in Supplement S4. If the MEL formulation, as proposed in the previous section for achiral objects, remains valid for embedded chiral objects; then certainly it will be a crucial verification in favor of the MEL method

Fig. 5 (a)-(d) shows the results of simulations using the chiral MEL stress tensor in the case of a 2D infinite cylinder embedded in a heterogeneous background. The mathematical consistency of this method is seen to be satisfied as seen in figures 5 (a)-(d). Consistency of the chiral MEL ST for other backgrounds (i.e. homogeneous and bounded backgrounds) is shown in detail in Supplement S4.

We may also note that the internal MEL stress tensor (and hence the internal bulk force) results in a correct time-averaged total force for several situations [3, 55], (or at least it follows the trend of the total time-averaged force). This is very useful from a computational point of view for both an embedded chiral and an achiral non-absorbing object.

## VI. MEL AND MCHU FORMULATIONS: ONLY MATHEMATICAL ALTERNATIVES OF THE MINKOWSKI'S FORCE LAW?

As discussed in Supplement S5, the important limitations of the MEL and MChu formulations include: (i) the use of a modified non-mechanical momentum density other than the Abraham or Minkowski density [details of the derivations given in APPENDIX B] and (ii) lack of relativistic invariance. In this regard, another important fact is that: even the well-known Abraham stress tensor and also the Einstein-Laub stress tensor [52] are not relativistically invariant [50,51]. Despite that, the Abraham stress tensor (but not the Einstein-Laub stress tensor) leads to the correct time-averaged total force for different real experiments [19,21]. As a result, considering all the relevant issues, it appears that Minkowski's force formulation is an appropriate description of the physical optical force. In spite of that, modified Chu (MChu) and modified Einstein-Laub (MEL) formulations undoubtedly lead to the correct time averaged total force on an embedded object (similar to Abraham ST). Note that the volumetric force of MChu formulation is widely applied [39] for force calculations.

As a result, if it would be possible to experimentally measure the presence of bulk force distribution of any ideal non-absorbing medium by any future experiment, it would certainly prove that even Minkowski's formulations are problematic. Such a future experiment on bulk force (especially in real-time analysis), if properly proceeds, will probably resolve many misconceptions and dilemmas in the area of distinct optical forces along with photon momenta. This may be possible by introducing modifications in the recent experiments of 'Ashkin-Dziedzic type' [25, 56-59]: by measuring the state of bulk part of water (or any ideal non-absorbing liquid) instead of the water surface.

## VII. CONCLUSION

This work explains the reasons behind the difference of results obtained using different time-averaged volumetric forces reported in [19,21] (e.g the Jones-Richard experiment, Jones-Leslie experiment, 1973 Ashkin-Dziedzic experiment, the experiment with a half immersed object, Hakim-Higham experiment, and so on). Our work also proposes possible effective solutions to the aforementioned problems, as follows:

(1) We require to modify the well-known Einstein-Laub and Chu formulations inside an embedded object to yield the correct time-averaged total force by Einstein-Laub and Chu formulations for all previous experiments like those involving a material background [2,3,19, 21, 31, 32, 37, 39-44].

The results of 'total time-averaged force' yielded by the modified EL and the modified Chu formulations are exactly equivalent to the 'total time-averaged force' given by Minkowski's and Abraham's forces and hence all the formulations can be made equivalent for 'total time-averaged real force' (even when material backgrounds are involved) as shown in Table A1 and A2 in APPENDIX A [though the volumetric force distribution process is fully different for them (as discussed: Minkowski force's total force arises fully from the surface of a lossless object)].

(2) Notably, all of our proposals remain valid not only for embedded achiral objects but also for chiral objects embedded in simple and complex backgrounds.

Furthermore, our work provides an efficient alternative solution for calculating the time-averaged bulk force with the modified Einstein-Laub (MEL) stress tensor method, saving much computational time and memory compared to those of the time-consuming bulk volumetric force method [39], both for embedded achiral [1-3,19, 21, 39-44, 55, 60-62] and chiral objects [31,32,37]. It should be noted that: background permittivity and permeability enter in the MEL stress tensor equation as ratio based unitless quantitities : $(\frac{\varepsilon_{b(j)}}{\varepsilon_s})$ and $(\frac{\mu_{b(j)}}{\mu_s})$. It should also be noted that: only the time averaged version of such a stress tensor should be considered for real world calculations.

After a detailed investigation (based on analytical calculations, full-wave simulations and real-world experiments), we have concluded that MEL and MChu formulations can certainly be used as very efficient mathematical toolkits to yield the time-averaged total force on an object embedded in a material background. If in future, any experiment (i.e. we have suggested one) detects the existence of bulk force, then it would require reconsidering the accepted Minkowski theory and demand examination of the MEL or MChu formulations.

**ACKNOWLEDGEMENTS**

M.R.C. Mahdy acknowledges for several important discussions: Associate Prof. Qiu Cheng Wei of National University of Singapore, Prof. Manuel Nieto Vesperinas of CSIC, Prof. Weiqiang Ding of Harbin Institute of Technology and Associate Prof. Dongliang Gao of Soochow University.

# APPENDIX A

**Table A1: Proposed modified time-averaged Einstein-Laub and Chu formulations**

| Name of Force | External ST | Time-averaged Internal ST | Time-averaged bulk force density | Surface force density | $\langle F_{Bulk} \rangle$ | $\langle F_{Surf} \rangle$ |
|---|---|---|---|---|---|---|
| Minkowski | $\bar{\bar{T}} = D_{out}E_{out} + B_{out}H_{out} - \frac{1}{2}\begin{pmatrix} B_{out} \cdot H_{out} \\ + D_{out} \cdot E_{out} \end{pmatrix}\bar{\bar{I}}$ | $\langle \bar{\bar{T}}_{Mink}^{out} \rangle = \begin{bmatrix} D_{in}E_{in}^* + B_{in}H_{in}^* \\ -\frac{1}{2}(B_{in} \cdot H_{in}^* + D_{in} \cdot E_{in}^*)\bar{\bar{I}} \end{bmatrix}$ | 0 | $f^{Surface} = -[\frac{1}{2}E_{out} \cdot E_{in}\Delta\varepsilon + \frac{1}{2}H_{out} \cdot H_{in}\Delta\mu]_{at\ r=a}$ | 0 | Non-zero (tot. force) |
| Modified Einstein-Laub (MEL) | Not Applicable | $\langle \bar{\bar{T}}_{MEL}^{Bulk} \rangle (in) = D_{in}E_{in}^* + B_{in}H_{in}^* - \frac{1}{2}\begin{pmatrix} (\frac{\mu_b}{\mu_s})\mu_s H_{in}^* \cdot H_{in} \\ +(\frac{\varepsilon_b}{\varepsilon_s})\varepsilon_s E_{in}^* \cdot E_{in} \end{pmatrix}\bar{\bar{I}}$ | $\langle f_{MEL}^{Bulk} \rangle(in) = (P_{Eff} \cdot \nabla)E_{in}^* + (M_{Eff} \cdot \nabla)H_{in}^* - (i\omega P_{Eff} \times B_{in}^*) + (i\omega M_{Eff} \times D_{in}^*)$. Here: $P_{Eff} = (\varepsilon_S - \varepsilon_b)E_{in}$; $M_{Eff} = (\mu_S - \mu_b)H_{in}$. | $f^{Surface} = \{\varepsilon_b(E_{out} - E_{in}) \cdot \hat{n}\} \left(\frac{E_{out} - E_{in}}{2}\right)_{at\ r=a} + \{\mu_b(H_{out} - H_{in}) \cdot \hat{n}\} \left(\frac{H_{out} - H_{in}}{2}\right)_{at\ r=a}$ | Non-zero | Non-zero |
| Modified Chu (MChu) | Not Applicable | $\langle \bar{\bar{T}}_{MCHU}^{Bulk} \rangle (in) = (\frac{\varepsilon_b}{\varepsilon_s})\varepsilon_s E_{in}E_{in}^* + (\frac{\mu_b}{\mu_s})\mu_s H_{in}H_{in}^* - \frac{1}{2}\begin{pmatrix} (\frac{\mu_b}{\mu_s})\mu_s H_{in}^* \cdot H_{in} \\ +(\frac{\varepsilon_b}{\varepsilon_s})\varepsilon_s E_{in}^* \cdot E_{in} \end{pmatrix}\bar{\bar{I}}$ | $\langle f_{MCHU}^{Bulk} \rangle(in) = -(\nabla \cdot P_{Eff})E_{in}^* - (\nabla \cdot M_{Eff})H_{in}^* - (i\omega P_{Eff} \times B_{in}^*) + (i\omega M_{Eff} \times D_{in}^*)$ | $f^{Surface} = [\{\varepsilon_b(E_{out} - E_{in}) \cdot \hat{n}\} \left(\frac{E_{out} + E_{in}}{2}\right)_{at\ r=a} + \{\mu_b(H_{out} - H_{in}) \cdot \hat{n}\} \left(\frac{H_{out} + H_{in}}{2}\right)_{at\ r=a}]$ | Non-zero | Non-zero |

Note: $\varepsilon_S$ and $\varepsilon_b$ are the permittivities of the scatterer (whose force is being calculated by employing the integration boundary enclosing it) and the background medium (in which that scatterer is embedded or immersed) respectively.

**Table A2: Satisfying the 'Consistency Conditions' for material backgrounds**

| Name of Force (x) | 'CII' [for material background]*** $\oint \langle \bar{\bar{T}}_{Mink}(out) \rangle \cdot ds = \langle F^{Total}(Consistent) \rangle = \langle F_x^{Bulk} \rangle(in) + \langle F_x^{Surface} \rangle$ satisfied? | 'CI' [for material background]*** $[\bar{\bar{T}}_{Mink}(out) - \bar{\bar{T}}_x(in)] \cdot \hat{n};\ _{at\ r=a} = f_x^{Surface}$ satisfied? |
|---|---|---|
| Minkowski | Yes | Yes |
| MEL | Yes | Yes |
| MChu* | Yes | Yes |

Notes: *(1) MChu = Modified Chu; and ***(2) $\bar{\bar{T}}$ (out) has been considered $\bar{\bar{T}}_{Mink}$ (out) due to three notable reasons. These 'notable' reasons have been discussed in the first sub-sect. of section III.

**APPENDIX B**

**A possible derivation of the internal Modified Einstein-Laub stress tensor of a generic object embedded in the material background**

According to ref. [39]:

"*In the frequency domain, and with Lumerical's sign convention of* $P(\omega) = \int e^{i\omega t} P(t) dt$, *we have* $i\omega P(\omega)$ *the final expression for the force per unit volume is therefore given by*

$$F_v = \varepsilon_0 (\nabla \cdot E) E + i\omega (P \times B).$$
$$= \varepsilon_0 (\nabla \cdot E) E + i\omega (\varepsilon_s - \varepsilon_0) E_{in} \times B.$$

*In a medium with a background index that is not 1, it is most numerically efficient and accurate to get the net optical force that will result in motion of the particle by rescaling the background permittivity, yielding a final equation*

$$F_v = \varepsilon_{br} \varepsilon_0 (\nabla \cdot E) E + i\omega \varepsilon_0 (\varepsilon_{sr} - \varepsilon_{br}) E_{in} \times B.$$

*where $\varepsilon_{br}$ is the background relative permittivity, and $\varepsilon_{sr}$ is the relative permittivity of the object.*

*Note that the equation for net force without rescaling with the background permittivity will give the total force on the volume including force on the background material which does not result in motion of the particle.*

*The volumetric technique is typically more accurate because many interpolation errors can be avoided. However, it can require a significant amount of memory because the electromagnetic fields and the permittivity must be recorded throughout the volume.*"

- In the last force equation [which is actually the modified version of Chu force applied previously in refs [26, 27, 60, 63, 64] [Also cf. Table A1 in APPENDIX A]. The term $(\varepsilon_S - \varepsilon_b) E_{in}$ can be defined as effective polarization, $P_{Eff} = (\varepsilon_S - \varepsilon_b) E_{in}$ [also cf. refs [26, 27, 44, 60, 63- 66] for such $P_{Eff}$], whereas the usual polarization $P = (\varepsilon_S - \varepsilon_0) E_{in}$. This simple change in the definition of well-known polarization is crucial for our forthcoming derivation of Modified Einstein-Laub stress tensor.

The well-known Einstein-Laub (EL) stress tensor was derived in [67] from the well-known EL force density method. In this section, we shall derive the modified EL ST from the modified EL (MEL) force in the same way as described in [67] but using the time-averaged form. The time-averaged form of well-known Einstein-Laub equation is written as [68]:

$$\langle \boldsymbol{f}_{EL} \rangle = \frac{1}{2}\text{Re}\left[(\boldsymbol{P}\cdot\nabla)\boldsymbol{E}_{in}^* - i\omega\left(\boldsymbol{P}\times\boldsymbol{B}_{in}^*\right)\right].$$

Similar to the modified Chu formulation (presented in [39] and in [26]), the EL force density takes a modified form for an embedded achiral dielectric object embedded in another homogeneous dielectric [27] (cf. Eq (8) given in ref [27]) and hence the correct time-averaged force on the particle can be obtained (excluding the force on the background according to [39]):

$$\langle \boldsymbol{f}_{MEL} \rangle = \frac{1}{2}\text{Re}\left[(\boldsymbol{P}_{Eff}\cdot\nabla)\boldsymbol{E}_{in}^* - i\omega\left(\boldsymbol{P}_{Eff}\times\boldsymbol{B}_{in}^*\right)\right]. \quad (B1)$$

In Eq. (B1), it is interesting to note that the polarization has been written as an effective polarization [26, 59, 62]:

$$\boldsymbol{D}_{in} = \boldsymbol{D}_{Eff} = \varepsilon_b \boldsymbol{E}_{in} + \boldsymbol{P}_{Eff}; \quad \boldsymbol{P}_{Eff} = (\varepsilon_S - \varepsilon_b)\boldsymbol{E}_{in}, \quad (B2)$$

In the same way, the effective magnetization can also be written as [26, 59]:

$$\boldsymbol{B}_{in} = \boldsymbol{B}_{Eff} = \mu_b \boldsymbol{H}_{in} + \boldsymbol{M}_{Eff}; \quad \boldsymbol{M}_{Eff} = (\mu_S - \mu_b)\boldsymbol{H}_{in}. \quad (B3)$$

Where $\varepsilon_b$ and $\mu_b$ are constant parameters. In [39] it is argued that: "*Note that the equation for net force without rescaling with the background permittivity will give the total force on the volume including force on the background material which does not result in motion of the particle.*" The idea of effective induced electric [65] (and magnetic [65]) dipole moment has also been applied previously to yield the 'outside force' of dipolar objects in [44,65,66] in a very similar way to the idea of effective polarization and magnetization given in Eq. (B2) and (B3). For example- the induced effective dipole moment has been derived in Eq. (22) in ref. [66] from the effective polarization model of Eq. (B2) but with the exterior field of an embedded scatterer for external dipolar force.

However, in [26], it is shown that for a magneto-dielectric object embedded in another magneto-dielectric medium, the external force calculated using the Minkowski stress tensor matches with the time-averaged force via modified Chu (MChu) volume force expressed in terms of effective polarization and magnetization as:

$$\langle \boldsymbol{f}_{Total}^{MCHU} \rangle = \langle \boldsymbol{f}_{Surface}^{MCHU} \rangle + \langle \boldsymbol{f}_{Bulk}^{MCHU} \rangle \quad (B4)$$

$$\langle \boldsymbol{f}_{Surface}^{MCHU} \rangle = \frac{1}{2}\text{Re}[\{\varepsilon_b(\boldsymbol{E}_{out} - \boldsymbol{E}_{in})\cdot\hat{\boldsymbol{n}}\}\left(\frac{\boldsymbol{E}_{out} + \boldsymbol{E}_{in}}{2}\right)^*_{at\ r=a} + \{\mu_b(\boldsymbol{H}_{out} - \boldsymbol{H}_{in})\cdot\hat{\boldsymbol{n}}\}\left(\frac{\boldsymbol{H}_{out} + \boldsymbol{H}_{in}}{2}\right)^*_{at\ r=a}] \quad (B5)$$

$$\left\langle \boldsymbol{f}_{\text{Bulk}}^{\text{MCHU}} \right\rangle = \frac{1}{2}\text{Re}[-i\omega\left(\boldsymbol{P}_{\text{Eff}} \times \boldsymbol{B}_{\text{in}}^*\right) + i\omega\left(\boldsymbol{M}_{\text{Eff}} \times \boldsymbol{D}_{\text{in}}^*\right)]. \qquad \text{(B6)}$$

For several cases [26, 27, 39, 59, 63, 64], this volume force method has been applied to yield the total force. For example- the effect of background permittivity has been described in [64] for the cases of binding forces. Considering the consistency of Eqs (B4-B6) [26,59], the modified EL force formulation in [27] for dielectric object embedded in another dielectric medium (cf. Eq (8) given in ref [27]) can be written in a generic form for a magneto-dielectric object embedded in another magneto-dielectric medium in accordance with [68] as:

$$\left\langle \boldsymbol{f}_{\text{MEL}}^{\text{Bulk}} \right\rangle (\text{in}) = \frac{1}{2}\text{Re}\left[ \left(\boldsymbol{P}_{\text{Eff}} \cdot \nabla\right)\boldsymbol{E}_{\text{in}}^* + \left(\boldsymbol{M}_{\text{Eff}} \cdot \nabla\right)\boldsymbol{H}_{\text{in}}^* - i\omega\boldsymbol{P}_{\text{Eff}} \times \left(\boldsymbol{B}_{\text{in}}^* - \boldsymbol{M}_{\text{Eff}}^*\right) + i\omega\boldsymbol{M}_{\text{Eff}} \times \left(\boldsymbol{D}_{\text{in}}^* - \boldsymbol{P}_{\text{Eff}}\right) \right]. \qquad \text{(B7)}$$

In Eq (B7), the $i\omega\boldsymbol{P}_{\text{Eff}} \times \boldsymbol{M}_{\text{Eff}}^*$ terms cancel out finally. As it is shown in [49] that the application of Einstein-Laub formulation may lead to misleading results for very simple situations, the final time-averaged form of MEL force should be written in terms of $\boldsymbol{B}$ instead of $\mu_0\boldsymbol{H}$ according to [68] (also cf. reference [69] where the difference between the time-varying and the time-averaged force has been argued):

$$\left\langle \boldsymbol{f}_{\text{MEL}}^{\text{Bulk}} \right\rangle (\text{in}) = \frac{1}{2}\text{Re}\left[ \left(\boldsymbol{P}_{\text{Eff}} \cdot \nabla\right)\boldsymbol{E}_{\text{in}}^* + \left(\boldsymbol{M}_{\text{Eff}} \cdot \nabla\right)\boldsymbol{H}_{\text{in}}^* - \left(i\omega\boldsymbol{P}_{\text{Eff}} \times \boldsymbol{B}_{\text{in}}^*\right) + \left(i\omega\boldsymbol{M}_{\text{Eff}} \times \boldsymbol{D}_{\text{in}}^*\right) \right]. \qquad \text{(B8)}$$

- *This is the final form of MEL volumetric force formulation for a magnetodielectric object embedded in another magneto dielectric. For the calculation of time-averaged optical total force on an object, above Eq (B8) should be used [not Eq (B7)]. The next part of this section will deal with the derivation of MEL ST.*

The well-known continuity equation of optical force is known as [18, 19, 21, 45]:

$$\nabla \cdot \overline{\overline{\boldsymbol{T}}} = \boldsymbol{f} + \frac{\partial}{\partial t}\boldsymbol{G}$$

Inside an object, time-averaged form of the above equation can be written as:

$$\left\langle \nabla \cdot \overline{\overline{\boldsymbol{T}}}(\text{in}) \right\rangle = \left\langle \boldsymbol{f}_{\text{MEL}}^{\text{Bulk}} \right\rangle (\text{in}) + \left\langle \frac{\partial}{\partial t}\boldsymbol{G}(\text{in}) \right\rangle$$

Already we have got the final form of $\left\langle \boldsymbol{f}_{\text{MEL}}^{\text{Bulk}} \right\rangle (\text{in})$ in Eq (B8). Now to find $\left\langle \nabla \cdot \overline{\overline{\boldsymbol{T}}} \right\rangle$, we need to figure out proper $\left\langle \frac{\partial}{\partial t}\boldsymbol{G} \right\rangle$.

Now, in order to yield the final form of the MEL stress tensor $\langle \bar{\bar{T}}_{MEL}^{Bulk} \rangle (in)$, let us consider the intermediate step of the MEL force [not the final time-averaged form given in Eq. (B8)] which can be written from Eq. (B7) as:

$$\langle f_{MEL}^{Bulk} \rangle (in) = \frac{1}{2} \text{Re}[(\boldsymbol{P}_{Eff} \cdot \nabla)\boldsymbol{E}_{in} + (\boldsymbol{M}_{Eff} \cdot \nabla)\boldsymbol{H}_{in} - i\omega \boldsymbol{P}_{Eff} \times \mu_b \boldsymbol{H}_{in} + i\omega \boldsymbol{M}_{Eff} \times \varepsilon_b \boldsymbol{E}_{in}] \quad (B9)$$

Again, Eq (B9), which represents only an intermediate step of final MEL force in Eq (B8), turns into the EL force equation if the background is air or vacuum. The only difference arises from the scaling of the air or vacuum permittivity and permeability by the relative permittivity and permeability of the background.

*It should be noted that the Maxwell equations remain unique everywhere in terms of **D** and **B***. However, as the internal induced polarization and magnetization are affected/influenced by the background permittivity and permeability [26, 27, 39, 44, 60, 63- 66] especially due to the effect of the boundary/interface, the division of **D** and **B** can be expressed based on Eq (B2) and (B3). As a result, considering $\boldsymbol{P}_{Eff}$ and $\boldsymbol{M}_{Eff}$ in Eq (B2) and (B3) as a consistent mathematical description (also applied previously in refs [26, 27, 39, 44, 60, 63- 66]); for time-harmonic fields, the Maxwell equations [70] can relate with them inside an embedded polarizable and magnetizable object as:

$$(\mu_b \boldsymbol{H}_{in} + \boldsymbol{M}_{Eff}) = \boldsymbol{B}_{in} = (\nabla \times \boldsymbol{E}_{in})/i\omega, \quad (B10)$$

$$(\varepsilon_b \boldsymbol{E}_{in} + \boldsymbol{P}_{Eff}) = \boldsymbol{D}_{in} = (\nabla \times \boldsymbol{H}_{in})/(-i\omega), \quad (B11)$$

$$\nabla \cdot (\varepsilon_b \boldsymbol{E}_{in} + \boldsymbol{P}_{Eff}) = \nabla \cdot \boldsymbol{D}_{in} = 0 \quad (B12)$$

$$\nabla \cdot (\mu_b \boldsymbol{H}_{in} + \boldsymbol{M}_{Eff}) = \nabla \cdot \boldsymbol{B}_{in} = 0 \quad (B13)$$

-These equations will be used in forthcoming steps.

We have stated earlier that: to determine the final form of the stress tensor, we need to consider the $\langle \frac{\partial}{\partial t} \boldsymbol{G} \rangle$ term properly. For time-harmonic fields, $\langle \frac{\partial}{\partial t} \boldsymbol{G} \rangle$ is zero where $\boldsymbol{G}$ is a non-mechanical momentum density. It appears that the usual relative permittivity and permeability are scaled by the background relative permittivity and permeability in MEL force density equation in Eq (B8) [also in its intermediate step given in Eq (B9)]. This is also true for the MChu force density given in [26, 39]. As a result, to determine the final form of internal MEL ST, the electromagnetic momentum density of Abraham ($\varepsilon_0 \mu_0 [\boldsymbol{E}_{in} \times \boldsymbol{H}_{in}]$ where $\varepsilon_0$ and $\mu_0$ are present), which is used to determine the well-known EL ST, can possibly be scaled by the relative $\varepsilon_b$ and relative $\mu_b$. Considering this fact, the appropriate $\boldsymbol{G}$ is considered for MEL formulation: $\boldsymbol{G}(in) = \varepsilon_b \mu_b [\boldsymbol{E}_{in} \times \boldsymbol{H}_{in}]$. In addition, $\text{Re}[\varepsilon_b \mu_b \boldsymbol{E}_{in} \times (-i\omega \boldsymbol{H}_{in}^*) - \varepsilon_b \mu_b \boldsymbol{H}_{in} \times (-i\omega \boldsymbol{E}_{in}^*)]$ term is zero, which is required to obtain $\langle \frac{\partial}{\partial t} \boldsymbol{G} \rangle = 0$ for time-harmonic fields. However, this $G(\text{in}) = \varepsilon_b \mu_b [\, E_{\text{in}} \times H_{\text{in}}\,]$ is neither Abraham not the Minkowski non-mechanical momentum density. In this regard, we prefer to indicate one of the conclusions in ref [71]: the non-mechanical momentum density inside matter may not be unique. For both MEL and MChu formulations, that conclusion in [71] is somehow quite appropriate.

As the $\text{Re}[\varepsilon_b \mu_b E_{\text{in}} \times (-i\omega H_{\text{in}}^*) - \varepsilon_b \mu_b H_{\text{in}} \times (-i\omega E_{\text{in}}^*)]$ term is zero, by adding this $\text{Re}[\varepsilon_b \mu_b E_{\text{in}} \times (-i\omega H_{\text{in}}^*) - \varepsilon_b \mu_b H_{\text{in}} \times (-i\omega E_{\text{in}}^*)]$ term with previous Eq. (B9), we finally obtain:

$$\left\langle \nabla \cdot \overline{\overline{T}} \right\rangle = \left\langle f_{\text{MEL}}^{\text{Bulk}} \right\rangle(\text{in}) + \left\langle \frac{\partial}{\partial t} G \right\rangle = \frac{1}{2}\text{Re}[[(P_{\text{Eff}} \cdot \nabla)E_{\text{in}}^* + (M_{\text{Eff}} \cdot \nabla)H_{\text{in}}^* - i\omega P_{\text{Eff}} \times \mu_b H_{\text{in}}^* + i\omega M_{\text{Eff}} \times \varepsilon_b E_{\text{in}}^*] \\ + [\varepsilon_b \mu_b E_{\text{in}} \times (-i\omega H_{\text{in}}^*) - \varepsilon_b \mu_b H_{\text{in}} \times (-i\omega E_{\text{in}}^*)]] \quad (B14)$$

After some calculations using Eqs (B10)- (B13), we get from Eq. (B14):

$$\left\langle \nabla \cdot \overline{\overline{T}} \right\rangle = \left\langle f_{\text{MEL}}^{\text{Bulk}} \right\rangle(\text{in}) = \frac{1}{2}\text{Re}[(P_{\text{Eff}} \cdot \nabla)E_{\text{in}}^* + (M_{\text{Eff}} \cdot \nabla)H_{\text{in}}^* - \varepsilon_b E_{\text{in}}^* \times (\nabla \times E_{\text{in}}) - \mu_b H_{\text{in}}^* \times (\nabla \times H_{\text{in}})] \quad (B15)$$

By applying Eqs (B2) and (B3), we can write from Eq (B15):

$$\left\langle \nabla \cdot \overline{\overline{T}} \right\rangle = \left\langle f_{\text{MEL}}^{\text{Bulk}} \right\rangle(\text{in}) = \frac{1}{2}\text{Re}[\nabla(D_{\text{in}} E_{\text{in}}^* + B_{\text{in}} H_{\text{in}}^*) - \frac{1}{2}\nabla(\varepsilon_b E_{\text{in}}^* \cdot E_{\text{in}}) - \frac{1}{2}\nabla(\mu_b H_{\text{in}}^* \cdot H_{\text{in}})] \quad (B16)$$

Hence the time-averaged MEL ST inside an object embedded in a homogeneous background can be written as:

$$\left\langle \overline{\overline{T}}_{\text{MEL}}^{\text{Bulk}} \right\rangle(\text{in}) = \frac{1}{2}\text{Re}[D_{\text{in}} E_{\text{in}}^* + B_{\text{in}} H_{\text{in}}^* - \frac{1}{2}\left((\frac{\mu_b}{\mu_s})\mu_s H_{\text{in}}^* \cdot H_{\text{in}} + (\frac{\varepsilon_b}{\varepsilon_s})\varepsilon_s E_{\text{in}}^* \cdot E_{\text{in}}\right)\overline{\overline{I}}]. \quad (B17)$$

Eq (B17) turns into EL Stress tensor [67] when the embedding background is air instead of a material medium. It should be noted that: background permittivity and permeability enter in the MEL stress tensor equation as ratio based unitless quantitities : $(\frac{\varepsilon_{b(j)}}{\varepsilon_s})$ and $(\frac{\mu_{b(j)}}{\mu_s})$. It should also be noted that: only the time averaged version of such a stress tensor should be considered for real world [72] calculations.

In the main article, we have discussed the consistency of Eq (B17) and Eq (B8) [bulk part of MEL force] based on our analytical calculation of time-averaged force on a magnetodielectric slab embedded in another magnetodielectric. However, still, one important issue remains unsolved: Are Eqs (B1) [8], (B8) and (B17) consistent equations, which lead to the same surface force from two different ways of calculations: (i) from direct volume force [38] and (ii) from the difference of external and internal ST. This issue is answered in the next section.

# APPENDIX C

**Derivation of the MEL surface force in two fully different ways**

At first, we consider the surface force calculation process shown in [38] (which applied in [38] to yield the surface force: (i) only from volumetric force and (ii) only for air background) and apply that process for the static part of Eq. (B8) [in APPENDIX B] at the boundary of the embedded object (at r=a) [considering there is no artificial gap between the scatterer and the background]:

$$\boldsymbol{f}_{\text{MEL}}^{\text{Surface}} = [\{\varepsilon_b(\boldsymbol{E}_{\text{out}} - \boldsymbol{E}_{\text{in}}) \cdot \hat{\boldsymbol{n}}\}\left(\frac{\boldsymbol{E}_{\text{out}} - \boldsymbol{E}_{\text{in}}}{2}\right)_{\text{at r=a}} + \{\mu_b(\boldsymbol{H}_{\text{out}} - \boldsymbol{H}_{\text{in}}) \cdot \hat{\boldsymbol{n}}\}\left(\frac{\boldsymbol{H}_{\text{out}} - \boldsymbol{H}_{\text{in}}}{2}\right)_{\text{at r=a}}]$$

(C1)

At the boundary of an object, the average polarization requires to be considered as $\boldsymbol{P}_{\text{Eff}}/2$ to arrive at Eq (C1) and Eq (C1) turns into the surface force of well-known Einstein-Laub force when the background is considered to be air. Now considering the difference between the external Minkowski ST [72,73] and the internal MEL ST just at the boundary of an embedded scatterer and applying the electromagnetic boundary conditions, we get:

$$[\bar{\bar{\boldsymbol{T}}}_{\text{Mink}}^{\text{MIX}}(\text{out}) - \bar{\bar{\boldsymbol{T}}}_{\text{MEL}}^{\text{MIX}}(\text{in})] \cdot \hat{\boldsymbol{n}}\Big|_{r=a} = [\varepsilon_b \boldsymbol{E}_{\text{out}}^{\perp} \cdot \boldsymbol{E}_{\text{out}}^{\perp} - \varepsilon_s \boldsymbol{E}_{\text{in}}^{\perp} \cdot \boldsymbol{E}_{\text{in}}^{\perp}]\hat{\boldsymbol{n}} + [\mu_b \boldsymbol{H}_{\text{out}}^{\perp} \cdot \boldsymbol{H}_{\text{out}}^{\perp} - \mu_s \boldsymbol{H}_{\text{in}}^{\perp} \cdot \boldsymbol{H}_{\text{in}}^{\perp}]\hat{\boldsymbol{n}}\Big|_{r=a}$$

(C2)

$$[\bar{\bar{\boldsymbol{T}}}_{\text{Mink}}^{\text{D}}(\text{out}) - \bar{\bar{\boldsymbol{T}}}_{\text{MEL}}^{\text{D}}(\text{in})] \cdot \hat{\boldsymbol{n}}\Big|_{r=a} = -\frac{1}{2}[[\varepsilon_b \boldsymbol{E}_{\text{out}}^{\perp} \cdot \boldsymbol{E}_{\text{out}}^{\perp} - \varepsilon_b \boldsymbol{E}_{\text{in}}^{\perp} \cdot \boldsymbol{E}_{\text{in}}^{\perp}]\hat{\boldsymbol{n}} + [\mu_b \boldsymbol{H}_{\text{out}}^{\perp} \cdot \boldsymbol{H}_{\text{out}}^{\perp} - \mu_b \boldsymbol{H}_{\text{in}}^{\perp} \cdot \boldsymbol{H}_{\text{in}}^{\perp}]\hat{\boldsymbol{n}}]\Big|_{r=a}$$

(C3)

Now, adding Eq (C2) and (C3) together, after some calculations we get:

$$[\bar{\bar{\boldsymbol{T}}}_{\text{Mink}}^{\text{out}}(\text{out}) - \bar{\bar{\boldsymbol{T}}}_{\text{MEL}}(\text{in})] \cdot \hat{\boldsymbol{n}}\Big|_{r=a} = [\{\varepsilon_b(\boldsymbol{E}_{\text{out}} - \boldsymbol{E}_{\text{in}}) \cdot \hat{\boldsymbol{n}}\}\left(\frac{\boldsymbol{E}_{\text{out}} - \boldsymbol{E}_{\text{in}}}{2}\right)_{\text{at r=a}} + \{\mu_b(\boldsymbol{H}_{\text{out}} - \boldsymbol{H}_{\text{in}}) \cdot \hat{\boldsymbol{n}}\}\left(\frac{\boldsymbol{H}_{\text{out}} - \boldsymbol{H}_{\text{in}}}{2}\right)_{\text{at r=a}}]$$

(C4)

-Which exactly matches with Eq. (C1).

**Figures and Captions**

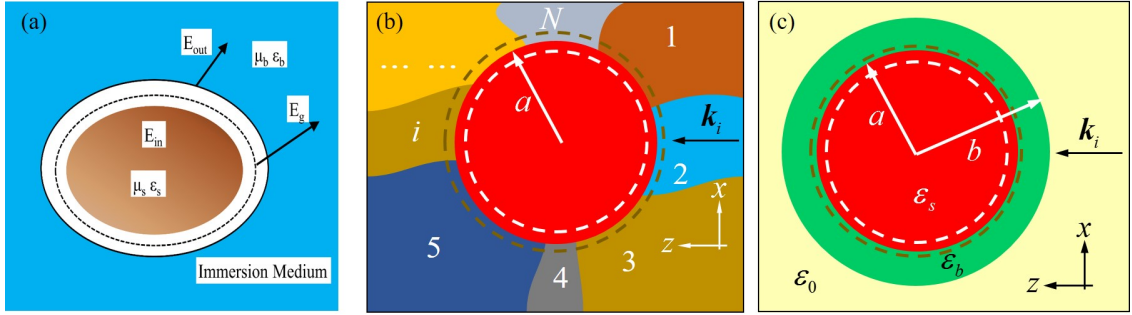

Fig. 1 Procedure of time-averaged optical force calculation by employing stress tensors. (a) '*gap method*': a small gap between the scatterer and the background should be considered. The time-averaged total force on the scatterer should be calculated using the time-averaged ST. $\langle \boldsymbol{F}_{\text{out}}^{\text{GAP}} \rangle$ evaluated *from fields strictly outside the object [i.e. gap field and background field]*, putting the integration boundary in the gap [20] (black dashed circle). However, the volume force calculation method is a little bit different, which is discussed in detail in [28]. (b) and (c) '*no gap method*': In both examples, the total force obtained by using the time-averaged ST is $\langle \boldsymbol{F}_{\text{out}} \rangle$ evaluated *from fields strictly outside the object considering no gap*, at $r = a^+ = 1.001a$, (black circles); whereas this force is $\langle \boldsymbol{F}_{\text{in}} \rangle$ or bulk force when the ST is determined *from fields strictly inside the object* considering no gap at $r = a^- = 0.999a$ (white circles). In (b), a sphere or cylinder is immersed in an unbounded and heterogeneous background. In (c) a core-shell sphere or cylinder (i.e., the core is embedded in a bounded background).

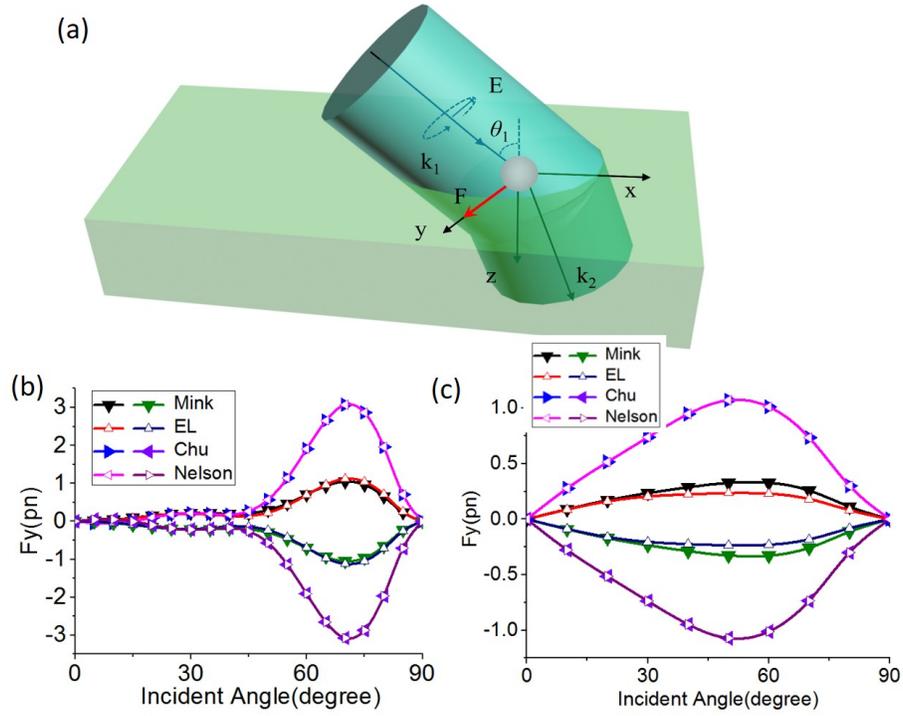

Fig. 2 (a) 3D Illumination geometry for a particle at the interface between two media to model the lateral force set-up in [12]. Either a right-handed or a left-handed circularly polarized (CP) beam of wavelength 1064 nm is incident along the $x$–$z$ plane at an angle $\theta_1$. $\mathbf{k_1}$ and $\mathbf{k_2}$ are wave vectors in media 1 and 2, respectively. The external force has been calculated based on the '*no gap method*'. (b) Transversal force (negative for a left-handed CP and positive for right-handed CP) as a function of angle of incidence for a 500 nm (radius) spherical $TiO_2$ particle located at the water-air interface calculated by external Minkowski, Einstein–Laub, Chu, and Amperian/Nelson ST. The Minkowski and Einstein-Laub STs predict a smaller lateral force in comparison with the time-averaged force yield by the external Chu and Ampere ST. (c) Transversal force, (negative for left-handed CP and positive for right-handed CP) as a function of the angle of incidence for an elliptical polystyrene (PS) particle [$r_x$=800 nm and $r_y$=$r_z$=(800/3) nm] located at the water-air interface, calculated by external Minkowski, Einstein–Laub, Chu, and Amperian STs. The Minkowski and Einstein-Laub STs predict lower lateral force in comparison with the time-averaged force yielded by the external Chu and Ampere STs.

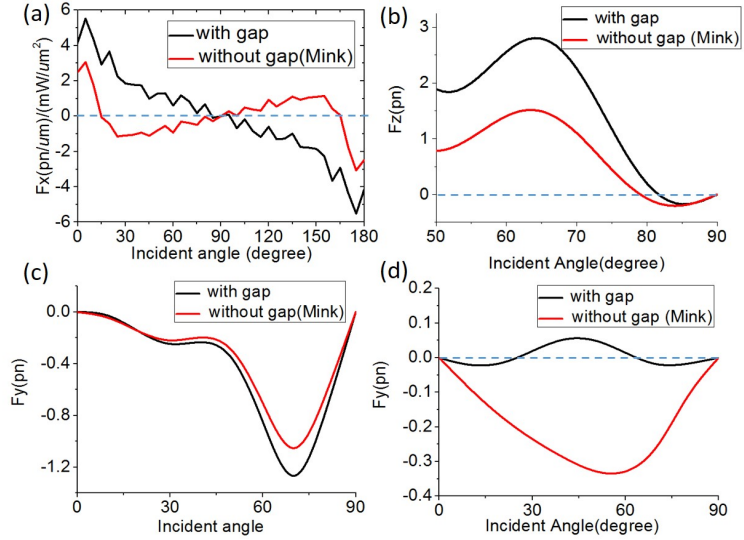

Fig. 3 Illustration of the optical force obtained from the '*gap method*'. (a) Optical force for a 2D spherical scatterer placed in an air-water interface, (cf. Fig. 2(d) and Fig. 2(e) in [3]). The parameters are $n_1$=1.0, $n_2$=1.33, $n_3$=1.45. The variation in the optical forces with the incidence angle [3] for the p-polarization case, as calculated via the external vacuum stress tensor considering a small gap of 2 nm, [the possible smallest gap with 2D full-wave simulation set-up; gap size << incident wavelength], between the scatterer and the water background. We have also examined our results by varying the size of the artificial gap, (i.e. 6nm, 10 nm and 20 nm). The results are almost same for all those gaps. The size of the scatterer is defined by $r_x = r_y$ =2.0 μm [3]. Instead of optical pulling [3], an optical pushing is achieved from the '*gap method*'. (b) Optical force (obtained by the vacuum ST) of a 3D dielectric particle given in Supplement S2, Fig. 2s (c), using two obliquely incident plane waves [1] *but considering a small gap of 10 nm* [possible smallest gap with a 3D full-wave simulation set-up; gap size << incident wavelength] between the water background and the embedded scatterer. Both '*gap method*' and '*no gap method*' lead to consistent results. (c) For the 3D set-up of Fig. 2(b), the time-averaged lateral force has been calculated from the vacuum ST considering a small gap (10nm) between the scatterer and the water background for a left-handed CP wave. (d) For the 3D set-up of Fig. 2(c), the time-averaged lateral force has been calculated from the vacuum ST considering a small gap (10 nm) between the scatterer and the water background for a left-handed CP wave. The sign of the lateral force is in disagreement with the time-averaged force yielded by the '*no gap method*' in Fig. 2(c).

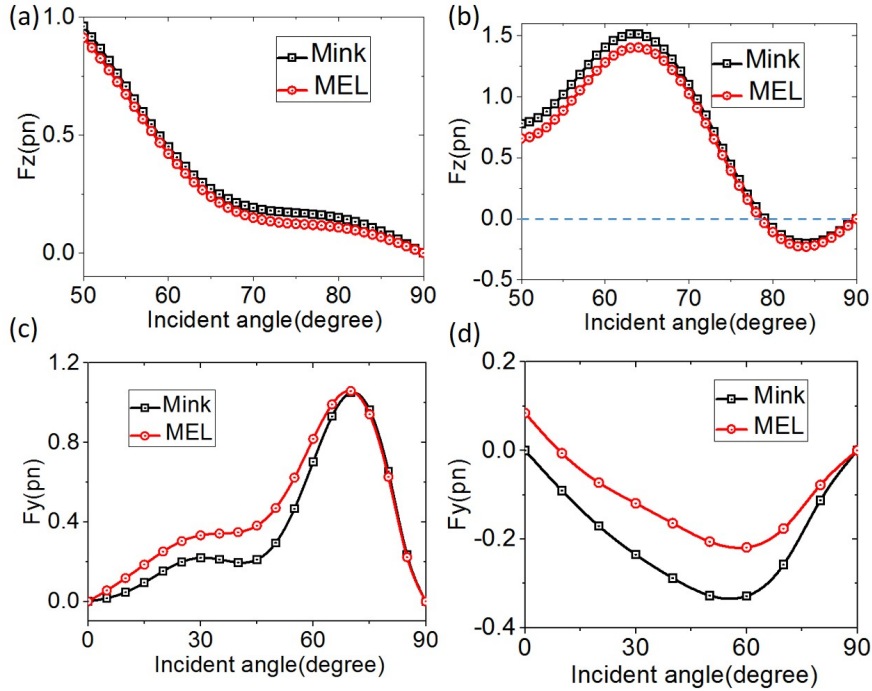

Fig. 4 Calculation of time-averaged total optical force ('*no gap method*') by external Minkowski ST and the time-averaged bulk force by internal MEL ST. These forces are always of same trend. By adding the surface force of achiral MEL with bulk force [cf. Eqs (6) and (7)], the magnitude exactly matches with external time-averaged total force by Minkowski ST. (a) For the two beamed tractor beam set-up in Supplement S2 Fig. 2s(b) with 320 nm object. (b) For the two beamed tractor beam set-up in Supplement S2 Fig. 2s(c) with 410 nm object. (c) For the lateral force set-up in Fig. 2(b) [only left-hand CP wave incident case] with spherical object. (d) For the lateral force set-up in Fig. 2(c) [only right-hand CP case] with elliptical object. For all the cases the trend of the time-averaged bulk force, obtained by employing the internal field only, is very similar to the total outside force calculated by the external Minkowski ST, using fields exterior to the scatterer. The bulk force by modified Chu volume force or stress tensor does not follow the trend of the total force.

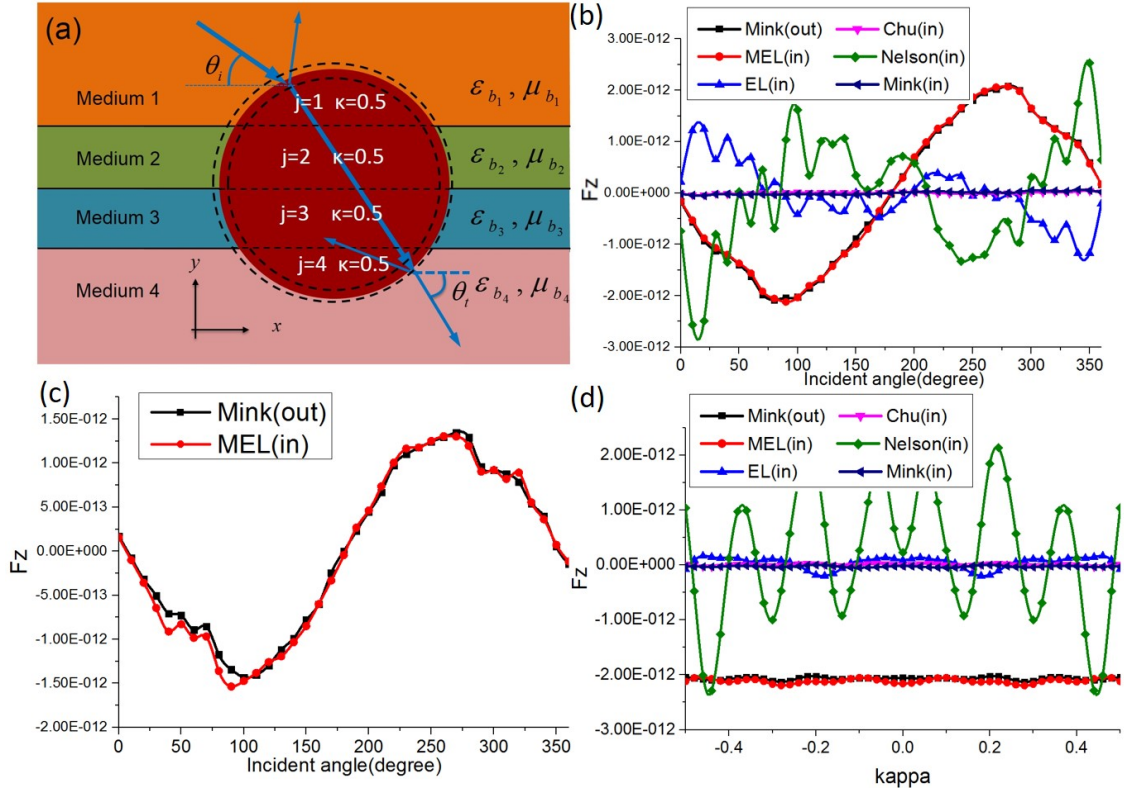

Fig. 5 Time-averaged forces: $F_{out}$ at $r = a^+ = 1.001a$ from Minkowski ST and $F_{in}$ (bulk force) at $r = a^- = 0.999a$ from the Chiral MEL ST. These forces have always the same trend. By adding the surface force of Chiral MEL with bulk force, the magnitude exactly matches with the external time-averaged total force. (a) Calculation procedure of force on a magneto-dielectric infinite chiral cylinder (chirality parameter, $\kappa = 0.5$) of $(\varepsilon_s, \mu_s) = (5\varepsilon_0, 4\mu_0)$ and radius 2000 nm, embedded in a heterogeneous unbounded background of four different magneto-dielectric layers: $(\varepsilon_b, \mu_b) = (3\varepsilon_0, 2\mu_0)$; $(4\varepsilon_0, 3\mu_0)$; $(5\varepsilon_0, 4\mu_0)$; $(6\varepsilon_0, 5\mu_0)$ at $\lambda = 1064$ nm. (b) Force on that cylinder when the plane wave $E_x = E_0 e^{i(kz - \omega t)}$ illuminates at varying angles of incidence. Notice that the internal force calculated by all other STs (i.e. EL, Chu, Nelson and Minkowski) has not in the same trend as the total external force. (c) Force on the same embedded cylinder when the illuminating circularly polarized wave: $E_x + iE_y : E_x = E_0 e^{i(kz - \omega t)} = E_y$ incidents at varying angles of incidence. (d) Force on the cylinder versus varying chiral parameter $\kappa$ when the illuminating plane wave incident at an angle of 45 degrees.

# Supplementary Information for "On the Modified Einstein-Laub and Modified Chu Optical Force Formulations"

M.R.C. Mahdy[1,2*], Tianhang Zhang[2], Golam Dastegir Al-Quaderi[3], Hamim Mahmud Rivy[1], Amin Kianinejad[2]

[1]*Department of Electrical & Computer Engineering, North South University, Bashundhara, Dhaka 1229, Bangladesh*

[2] *Department of Electrical and Computer Engineering, National University of Singapore, 4 Engineering Drive 3, Singapore 117583*

[3] *Department of Physics, University of Dhaka, Dhaka-1000, Bangladesh*

[*] Corresponding author: mahdy.chowdhury@northsouth.edu

## S1. Surface force of Chu method and a detailed analysis on *'gap method'* of optical force calculation:

We discuss the surface force determination of Chu force. It is shown in [1] that when we consider an object placed in air, the surface force of Chu can be determined from the volumetric force of Chu by appropriate boundary conditions [1-3]:

$$\boldsymbol{f}_{\text{Chu}}^{\text{Surface}} = \{\varepsilon_0 (\boldsymbol{E}_{\text{out}} - \boldsymbol{E}_{\text{in}}) \cdot \hat{\boldsymbol{n}}\} \left(\frac{\boldsymbol{E}_{\text{out}} + \boldsymbol{E}_{\text{in}}}{2}\right)_{\text{at } r=a} + \{\mu_0 (\boldsymbol{H}_{\text{out}} - \boldsymbol{H}_{\text{in}}) \cdot \hat{\boldsymbol{n}}\} \left(\frac{\boldsymbol{H}_{\text{out}} + \boldsymbol{H}_{\text{in}}}{2}\right)_{\text{at } r=a} \quad (1s)$$

However, in [1] the surface force has not been calculated by employing the stress tensors. We shall show the process of surface force calculation by employing the stress tensor method in this section. The internal Chu stress tensor (ST) can be written as [4]:

$$\bar{\bar{T}}_{\text{Chu}}^{\text{in}} = \left[\varepsilon_0 \boldsymbol{E}_{\text{in}} \boldsymbol{E}_{\text{in}}^* + \mu_0 \boldsymbol{H}_{\text{in}} \boldsymbol{H}_{\text{in}}^* - \frac{1}{2}\left(\mu_0 \boldsymbol{H}_{\text{in}} \cdot \boldsymbol{H}_{\text{in}}^* + \varepsilon_0 \boldsymbol{E}_{\text{in}} \cdot \boldsymbol{E}_{\text{in}}^*\right)\bar{\bar{I}}\right].$$ Here 'in' represents the fields inside a

scatterer. Now the surface force of Chu in Eq. (1s) can be yielded from the difference of external vacuum stress tensor $\bar{\bar{T}}_{\text{Vacuum}}$ (out) and the internal Chu ST [cf. Eq. (2b) in the main article for vacuum ST; for the case of air background: the gap field ($\boldsymbol{E}_g$) itself is the outside field ($\boldsymbol{E}_{\text{out}}$) in Fig. 1s]:

$$[\bar{\bar{T}}_{\text{Vacuum}}^{\text{MIX}}(\text{out}) - \bar{\bar{T}}_{\text{Chu}}^{\text{MIX}}(\text{in})] \cdot \hat{\boldsymbol{n}} \Big|_{r=a} = [\varepsilon_0 E_{\text{out}}^{\perp} \cdot E_{\text{out}}^{\perp} - \varepsilon_0 E_{\text{in}}^{\perp} \cdot E_{\text{in}}^{\perp}]\hat{\boldsymbol{n}} + [\varepsilon_0 E_{\text{m out}}^{\parallel} \cdot E_{\text{out}}^{\perp} - \varepsilon_0 E_{\text{m in}}^{\parallel} \cdot E_{\text{in}}^{\perp}]\hat{\boldsymbol{m}}$$
$$+ [\varepsilon_0 E_{\text{q out}}^{\parallel} \cdot E_{\text{out}}^{\perp} - \varepsilon_0 E_{\text{q in}}^{\parallel} \cdot E_{\text{in}}^{\perp}]\hat{\boldsymbol{q}}$$
$$+ [\mu_0 H_{\text{out}}^{\perp} \cdot H_{\text{out}}^{\perp} - \mu_0 H_{\text{in}}^{\perp} \cdot H_{\text{in}}^{\perp}]\hat{\boldsymbol{n}} + [\mu_0 H_{\text{m out}}^{\parallel} \cdot H_{\text{out}}^{\perp} - \mu_0 H_{\text{m in}}^{\parallel} \cdot H_{\text{in}}^{\perp}]\hat{\boldsymbol{m}}$$
$$+ [\mu_0 H_{\text{q out}}^{\parallel} \cdot H_{\text{out}}^{\perp} - \mu_0 H_{\text{q in}}^{\parallel} \cdot H_{\text{in}}^{\perp}]\hat{\boldsymbol{q}}$$

'MIX' represents the mixed diagonal and non-diagonal elements of the stress tensor, which are not connected with the identity tensor, $\bar{\bar{I}}$. 'out' represents the total fields (incident plus scattered field) outside a scatterer. Electric field at the object and background boundary is defined as: $\boldsymbol{E} = E^{\perp}\hat{\boldsymbol{n}} + E_{\text{m}}^{\parallel} \hat{\boldsymbol{m}} + E_{\text{q}}^{\parallel} \hat{\boldsymbol{q}}$ where $\hat{\boldsymbol{n}}, \hat{\boldsymbol{m}}$ and $\hat{\boldsymbol{q}}$ are mutually orthogonal arbitrary unit vectors, which are applicable for different coordinate systems such as Cartesian or Spherical or Cylindrical. $\hat{\boldsymbol{n}}$ is the local outward unit normal of the object surface. $E^{\parallel}$ and $E^{\perp}$ are the parallel and perpendicular components of electric fields at the background and object boundary. In a very similar way, the magnetic field can also be defined. Now, by applying the electromagnetic boundary conditions, we finally arrive for mix components:

$$[\bar{\bar{T}}_{\text{Vacuum}}^{\text{MIX}}(\text{out}) - \bar{\bar{T}}_{\text{Chu}}^{\text{MIX}}(\text{in})] \cdot \hat{n}\Big|_{r=a} = \begin{bmatrix} [\varepsilon_0 E_{\text{out}}^{\perp} \cdot E_{\text{out}}^{\perp} - \varepsilon_0 E_{\text{in}}^{\perp} \cdot E_{\text{in}}^{\perp}]\hat{n} + \dfrac{\varepsilon_0}{2}\left(E_{\text{m out}}^{\parallel} + E_{\text{m in}}^{\parallel}\right)[E_{\text{out}}^{\perp} - E_{\text{in}}^{\perp}]\hat{m} \\ + \dfrac{\varepsilon_0}{2}\left(E_{\text{q out}}^{\parallel} + E_{\text{q in}}^{\parallel}\right)[E_{\text{out}}^{\perp} - E_{\text{in}}^{\perp}]\hat{q} \end{bmatrix}_{r=a}$$

$$+ \begin{bmatrix} [\mu_0 H_{\text{out}}^{\perp} \cdot H_{\text{out}}^{\perp} - \mu_0 H_{\text{in}}^{\perp} \cdot H_{\text{in}}^{\perp}]\hat{n} + \dfrac{\mu_0}{2}\left(H_{\text{m out}}^{\parallel} + H_{\text{m in}}^{\parallel}\right)[H_{\text{out}}^{\perp} - H_{\text{in}}^{\perp}]\hat{m} \\ + \dfrac{\mu_0}{2}\left(H_{\text{q out}}^{\parallel} + H_{\text{q in}}^{\parallel}\right)[H_{\text{out}}^{\perp} - H_{\text{in}}^{\perp}]\hat{q} \end{bmatrix}_{r=a}$$

(2s)

Now, we can also write for pure diagonal components (those are connected with $\bar{\bar{I}}$) of stress tensor:

$$[\bar{\bar{T}}_{\text{Vacuum}}^{\text{D}}(\text{out}) - \bar{\bar{T}}_{\text{Chu}}^{\text{D}}(\text{in})] \cdot \hat{n}\Big|_{r=a} = -\frac{1}{2}[[\varepsilon_0 E_{\text{out}}^{\perp} \cdot E_{\text{out}}^{\perp} - \varepsilon_0 E_{\text{in}}^{\perp} \cdot E_{\text{in}}^{\perp}]\hat{n} + [\mu_0 H_{\text{out}}^{\perp} \cdot H_{\text{out}}^{\perp} - \mu_0 H_{\text{in}}^{\perp} \cdot H_{\text{in}}^{\perp}]\hat{n}]_{r=a}$$

(3s)

'D' represents pure diagonal elements of the stress tensor. Now, by adding Eqs (2s) and (3s), after some calculations we get exactly Eq (1s), the surface force of Chu:

$$[\bar{\bar{T}}_{\text{Vacuum}}(\text{out}) - \bar{\bar{T}}_{\text{Chu}}(\text{in})] \cdot \hat{n}\Big|_{r=a} = [\{\varepsilon_0(E_{\text{out}} - E_{\text{in}}) \cdot \hat{n}\}\left(\dfrac{E_{\text{out}} + E_{\text{in}}}{2}\right)_{\text{at } r=a} + \{\mu_0(H_{\text{out}} - H_{\text{in}}) \cdot \hat{n}\}\left(\dfrac{H_{\text{out}} + H_{\text{in}}}{2}\right)_{\text{at } r=a}]$$

(4s)

So, when the object is placed in air: (i) internal Chu stress tensor yields the time-averaged bulk force of Chu volume force and (ii) the external vacuum ST leads to the total time-averaged force (bulk force plus the additional surface force). This is also true for the Einstein-Laub volumetric force. We have calculated that: at the object boundary, the difference of external vacuum ST and the internal Einstein-Laub ST leads to a surface force of Einstein-Laub, which exactly matches with the independently calculated surface force from the volumetric force of Einstein-Laub. It should be noted that in [1] the surface force has been calculated only from the volumetric force approach.

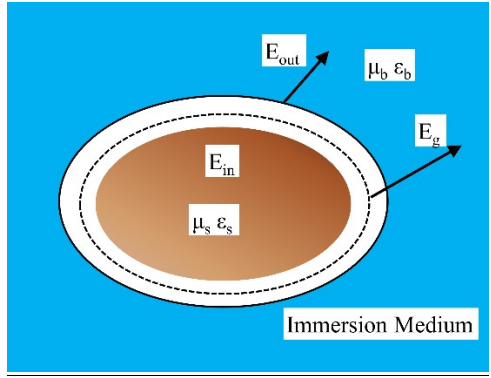

Fig. 1s: When the object is surrounded by a medium other than the free space, a narrow gap may be imagined to exist between the object and its surroundings according to refs. [2-5]. The integration boundary is then placed within the gap. $E_g$ represents the electric field in that narrow gap.

We now discuss the volume force method of *'gap method'* [2-5] (defined in the introduction of the main article and also cf. Fig. 1s), In [3] considering the consistency of the volume force, it has been concluded that when such a gap is given between the scatterer and the surrounding background, the total force of the embedded scatterer should be calculated based on (described as method II in [3] and cf. Fig. 1s):

$$\left\langle F_{\text{Gap Method}}^{\text{Total}} \right\rangle = \left\langle F_{\text{Gap Method}}^{\text{Bulk}} \right\rangle(\text{in}) + \left\langle F_{\text{Gap Method}}^{\text{Surface}} \right\rangle \tag{5as}$$

$$\left\langle f_{\text{Gap Method}}^{\text{Surface}} \right\rangle = \frac{1}{2}\text{Re}[\{\varepsilon_0 (E_g - E_{\text{in}}) \cdot \hat{n}\} \left( \frac{E_{\text{out}}^{\perp} + E_{\text{in}}^{\perp}}{2} \right)_{\text{at } r=a}^{*} + \{\mu_0 (H_g - H_{\text{in}}) \cdot \hat{n}\} \left( \frac{H_{\text{out}}^{\perp} + H_{\text{in}}^{\perp}}{2} \right)_{\text{at } r=a}^{*} ] \tag{5bs}$$

$$\left\langle f_{\text{Gap Method}}^{\text{Bulk}} \right\rangle(\text{in}) = \frac{1}{2}\text{Re}\left[ -(i\omega P \times B_{\text{in}}^{*}) + (i\omega M \times D_{\text{in}}^{*}) \right]. \tag{5cs}$$

Here $P = (\varepsilon_S - \varepsilon_0) E_{\text{in}}$, $M = (\mu_S - \mu_0) H_{\text{in}}$, $B_{\text{in}} = \mu_S H_{\text{in}}$ and $D_{\text{in}} = \varepsilon_S E_{\text{in}}$. But the bulk and surface force calculation based on the above method may not be a general procedure due to these reasons:

(1) The surface force of Chu force calculated in [1] from the direct volume force method. It is also shown in Eqs. (1s-4s) that the same surface force can also be calculated for a dielectric object (placed in air) from the difference of the external ST, $\bar{\bar{T}}_{Vacuum}(out)$, and the internal ST of Chu, $\bar{\bar{T}}_{Chu}(in)$, just at the boundary of the scatterer. So, the total force of a scatterer placed in air calculated by volume force method is nothing but the force calculated by the external ST.

Now, we again consider the work of Barnett et al. [1] where the procedure of surface force calculation has been illustrated by properly applying the boundary conditions. If we apply those boundary conditions and at the same time consider that there is a small gap between the scatterer and the surrounding background, the consistent equation leads to the force equation where the average field at the boundary of the scatterer should be calculated with gap fields ($E_g, H_g$) instead of the real fields ($E_{out}, H_{out}$) of the background medium (described as method III in [3] and cf. Fig. 1s):

$$\langle f^{Surface}_{Gap\ Method} \rangle = \frac{1}{2} \text{Re}[\{\varepsilon_0 (E_g - E_{in}) \cdot \hat{n}\} \left(\frac{E_g^\perp + E_{in}^\perp}{2}\right)^*_{at\ r=a} + \{\mu_0 (H_g - H_{in}) \cdot \hat{n}\} \left(\frac{H_g^\perp + H_{in}^\perp}{2}\right)^*_{at\ r=a}] \quad (6as)$$

It is also interesting that if we consider a vacuum stress tensor at the gap [4] shown between the scatterer and the background, the difference between the external vacuum stress tensor and the internal Chu stress tensor leads to the same surface force of Eq (6as):

$$[\bar{\bar{T}}^{GAP}_{Vacuum}(out) - \bar{\bar{T}}_{Chu}(in)] \cdot \hat{n}\Big|_{r=a} = [\{\varepsilon_0 (E_g - E_{in}) \cdot \hat{n}\} \left(\frac{E_g^\perp + E_{in}^\perp}{2}\right)_{at\ r=a} + \{\mu_0 (H_g - H_{in}) \cdot \hat{n}\} \left(\frac{H_g^\perp + H_{in}^\perp}{2}\right)_{at\ r=a}]$$

(6bs)

But that equation (method III in [3]) does not lead to the consistent result according to that same ref. [3]. In addition, we have already shown in the main article that the total force calculated by external ST $\bar{\bar{T}}^{GAP}_{Vacuum}(out)$ in Eq (6bs) [method III in [3] should also lead to the same force of Eq (2a) given in the main article] leads to wrong result for experiments which include inhomogeneous background (or in general heterogeneous background).

So, Eq (5as) above is an independent equation (described as method II in [3] which is a more accurate formulation than method I and III according to ref. [3]) which cannot be derived from the volume force of Chu [1] by employing the proper boundary conditions to the static force part shown in [1]. At the same time, there is no such external (and also internal)

ST which leads to the surface force yielded by Eq (5bs) above from the difference of external ST and the internal Chu ST just at the object boundary.

(2) Last but not the least, the difference of any external ST and the internal Chu ST does not lead to the surface force given in Eq (5bs). So, Eq (5as) above (described as method II in [3]) is a fully independent equation to yield the time-averaged total force, which may not be a generic way to yield the total time-averaged force of the embedded objects. Though above Eq. (5as) is successful for small objects only for a few specific cases [3], according to the same ref. [3] it may lead to inconsistent results for embedded Mie or more complex objects (we believe: due to aforementioned reasons discussed here).

An alternative way to yield the time-averaged total force (by volume force method) for the embedded objects can be well-known Einstein-Laub force [6, 7] as shown for several previous experiments [6] but without considering any small gap between the scatterer and the background. However, the problems of EL force [7] for real experiments, those involve material background, have been discussed shortly in the main article.

**S2. Two-beamed tractor beam experiment: Calculation of total force based on different external stress tensors (STs)**:

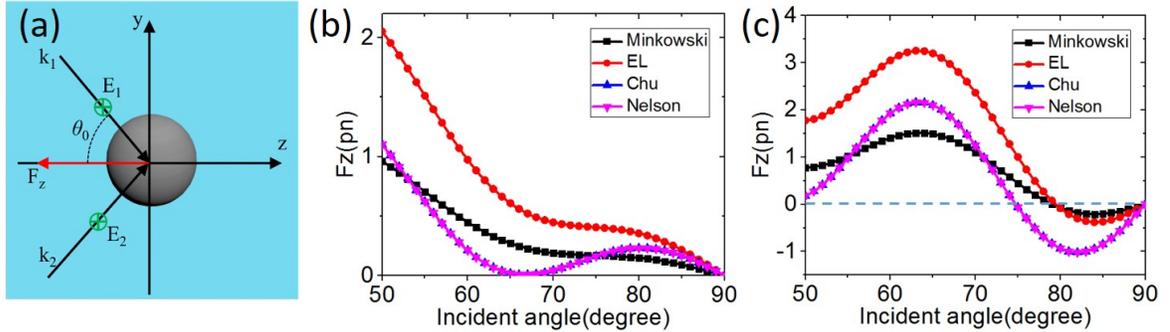

Fig. 2s: (a) Optical sorting of a dielectric particle using two obliquely incident plane waves reported in [8]. The external force will be calculated based on 'no gap method'. Two TE-polarized plane waves ($\lambda$ = 532 nm) incident at *varying angle* $\theta$ onto a polystyrene cylinder ($n$ = 1.58) immersed in water ($n$ = 1.33), shown: (b) For $r$ = 320 nm, at a steady state when the plane waves exert a pushing force in the +$x$ direction. Force on the bead calculated by external Minkowski, Einstein–Laub, Chu, and Ampere/Nelson ST for TE polarization as a function of incident angle $\theta$. (c) For $r$ = 410 nm, at a steady state when the plane waves exert a pulling force in the −$x$ direction. Force on the bead calculated by external Minkowski, Einstein–Laub, Chu, and Ampere/Nelson ST for TE polarization as a function of incident angle $\theta$. Pulling forces are only achieved for TE polarization within a short range of angles experimentally observed in [8].

## S3. Time averaged external (total force) and internal force (bulk force) for an achiral object embedded in complex backgrounds:

All the 3D simulations throughout the main article and this supplement are conducted using incident power of $0.57 \text{mW}/\mu\text{m}^2$.

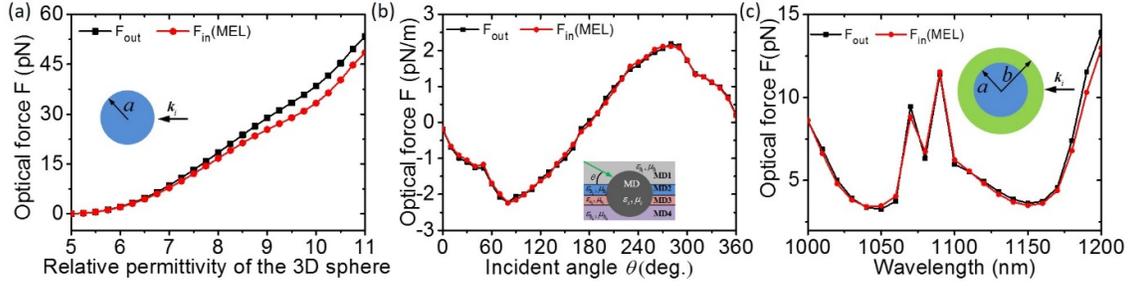

Fig. 3s: Time-averaged forces: $F_{out}$ at $r = a^+ = 1.001a$ from Minkowski ST and $F_{in}$ (bulk force) at $r = a^- = 0.999a$ from the MEL ST. These forces are always of same trend. By adding the surface force of MEL with the bulk force, the magnitude exactly matches with external time-averaged total force. (a) Homogeneous background: Force on a non-absorbing dielectric sphere by varying its permittivity (with $a$=500 nm, illuminated by a linearly polarized plane wave $E_x = E_0 e^{i(kz-\omega t)}$ at $\lambda = 1064$ nm). The unbounded homogeneous dielectric background parameter is: $\varepsilon_b = 5\varepsilon_0$. (b) Heterogeneous background: Force on a magneto-dielectric infinite cylinder of $(\varepsilon_s, \mu_s) = (5\varepsilon_0, 4\mu_0)$ and radius 2000 nm embedded in a heterogeneous unbounded background of four different magneto-dielectric layers: $(\varepsilon_b, \mu_b) = (3\varepsilon_0, 2\mu_0); (4\varepsilon_0, 3\mu_0); (5\varepsilon_0, 4\mu_0); (6\varepsilon_0, 5\mu_0)$. The illuminating plane wave with $\lambda = 1064$ nm indices at varying angles. (c) A 3D core-shell magneto-dielectric sphere embedded in air is illuminated by a linearly polarized plane wave with varying wavelengths. Core radius, $a$=500 nm, $\varepsilon_s = 8\varepsilon_0$, $\mu_s = 3\mu_0$. Bounded local immediate background (i.e. the shell) parameters: radius, $b$=600 nm and $\varepsilon_b = 4\varepsilon_0$; $\mu_s = 2\mu_0$. $\boldsymbol{F}_{out} = \langle \boldsymbol{F}_{out}^{Core} \rangle$, at different illumination wavelengths $\lambda$ obtained from Minkowski ST at $r=a^+$ using the fields in the shell. $\boldsymbol{F}_{in} = \langle \boldsymbol{F}_{in}^{Core} \rangle$ based on MEL ST at $r=a^-$ using core fields.

## S4. Time averaged external (total force) and internal force (bulk force) for a chiral object embedded in material background:

In this section we shall also show the consistency of the proposed internal MEL method with the external Minkowski ST method [9] for embedded chiral objects considering both simple and complex backgrounds. If a magneto-dielectric chiral object is embedded in a material background, the proposed internal MEL stress tensor:

$$\left\langle \overline{\overline{T}}_{MEL(j)}^{chiral} \right\rangle (in) = D_{in}^{chiral} E_{in}^{*} + B_{in}^{chiral} H_{in}^{*} - \frac{1}{2}\left( (\frac{\mu_{b(j)}}{\mu_s})\mu_s H_{in}^{*} \cdot H_{in} + (\frac{\varepsilon_{b(j)}}{\varepsilon_s})\varepsilon_s E_{in}^{*} \cdot E_{in} \right) \overline{\overline{I}}. \quad (7s)$$

Where $D_{in}^{chiral} = \varepsilon_s E_{in} - j(\kappa/c)H_{in}$; $B_{in}^{chiral} = \mu_s H_{in} + j(\kappa/c)E_{in}$. The total time-averaged bulk force on the embedded object should be: $\left\langle F_{Bulk} \right\rangle (in) = \sum_j \iint \left\langle \overline{\overline{T}}_{MEL(j)}^{chiral} \right\rangle (in) \cdot ds_{(j)}$. The Bulk force of chiral MEL method can also be written similar to the achiral MEL volumetric force density [Eq (4) in the main article] as:

$$\left\langle f_{MEL(j)}^{Chiral} \right\rangle (\text{Bulk}) = \frac{1}{2}\text{Re}\left[ \left( P_{Chiral(j)} \cdot \nabla \right) E_{in}^{*} + \left( M_{Chiral(j)} \cdot \nabla \right) H_{in}^{*} - \left( i\omega P_{Chiral(j)} \times B_{in}^{*} \right) + \left( i\omega M_{Chiral(j)} \times D_{in}^{*} \right) \right]. \quad (8s)$$

In Eq. (8s), the effective polarization and magnetization are defined as: $P_{chiral} = P_e + M_c$, $P_e = (\varepsilon_S - \varepsilon_b)E_{in}$, $M_c = -j(\kappa/c)H_{in}$ and $M_{chiral} = M_n + P_c$, $M_n = (\mu_S - \mu_b)H_{in}$, $P_c = j(\kappa/c)E_{in}$ respectively. However, $D_{in}^{chiral}$ and $B_{in}^{chiral}$ in Eq. (8s) should be written directly as: $D_{in}^{chiral} = \varepsilon_s E_{in} - j(\kappa/c)H_{in}$; $B_{in}^{chiral} = \mu_s H_{in} + j(\kappa/c)E_{in}$. The surface force part of the total force can also be derived similarly to that of the achiral case. The total time-averaged force should be the surface force plus the bulk force.

Consistency of chiral MEL ST has been shown for chiral objects embedded in homogeneous (in Fig. 4s) and in bounded (in Fig. 5s) background medium. In the main article, we have also discussed the case of heterogeneous background.

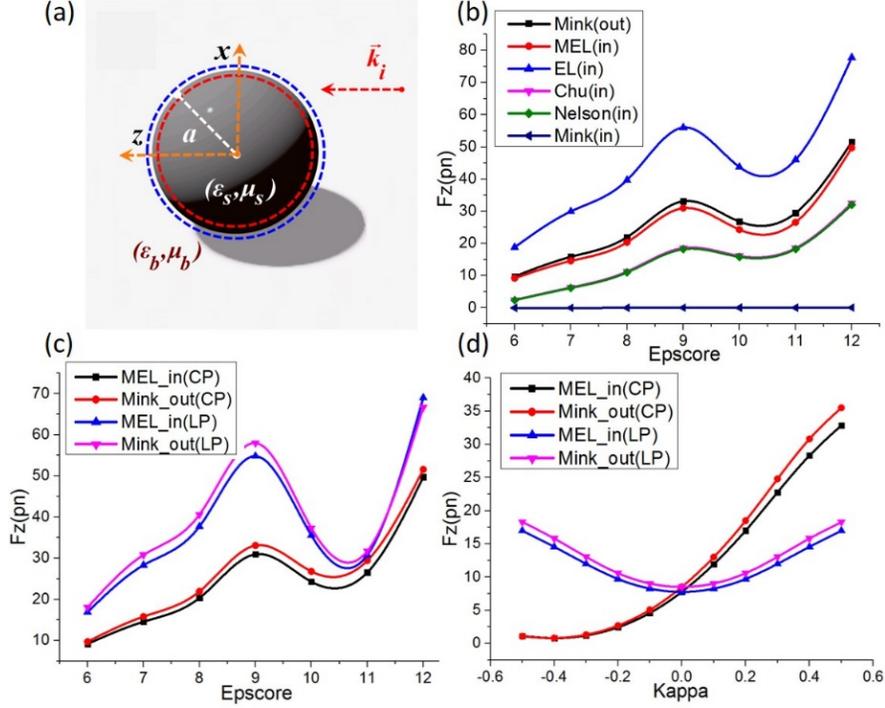

FIG.4s: Time-averaged forces: $F_{out}$ at $r = a^+ = 1.001a$ from Minkowski ST and $F_{in}$ (bulk force) at $r = a^- = 0.999a$ from the Chiral MEL ST. These forces are always of same trend. By adding the surface force of Chiral MEL with bulk force, the magnitude exactly matches with external time-averaged total force. (a) Calculation procedure of force on a dielectric sphere with $a$=500 nm (i.e., a Mie object) at $\lambda = 1064$ nm. The unbounded homogeneous dielectric background parameter: $\varepsilon_b = 4\varepsilon_0$. (b) Force on that chiral dielectric sphere (chirality parameter, $\kappa = 0.4$) by varying the permittivity of the sphere, illuminated by a linearly polarized plane wave $E_x = E_0 e^{i(kz-\omega t)}$. Notice that the internal forces (bulk forces) calculated by all other STs (i.e. EL, Chu, Nelson and Minkowski) are not so close to the total time-averaged force. (c) Force on the same embedded chiral dielectric sphere by varying the permittivity of the sphere, illuminated by a linear polarized and a circularly polarized ($E_x + iE_y : E_x = E_0 e^{i(kz-\omega t)} = E_y$) wave. (d) Force on the embedded chiral dielectric ($\varepsilon_s = 4\varepsilon_0$) sphere by varying the chirality parameter of the sphere, illuminated by a linear polarized and a circularly polarized wave.

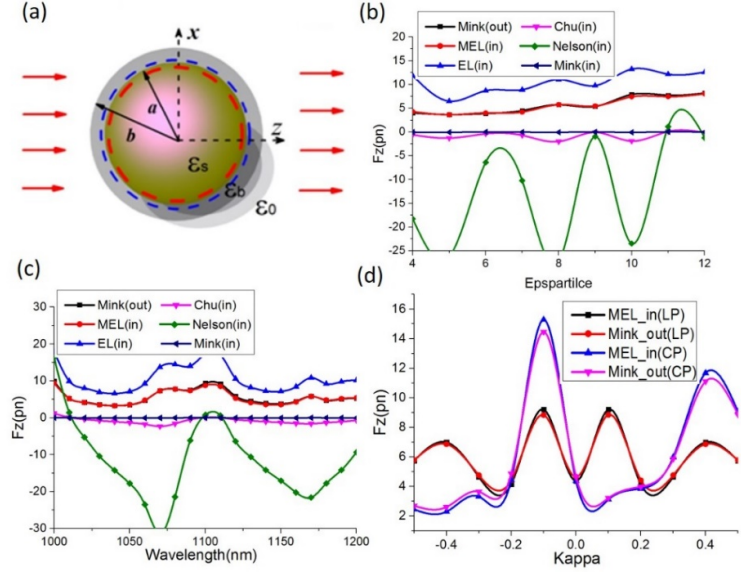

FIG. 5s: Time-averaged forces: $F_{out}$ at $r=a^+ =1.001a$ from Minkowski ST and $F_{in}$ (bulk force) at $r=a^- =0.999a$ from the Chiral MEL ST. These forces are always of same trend. By adding the surface force of Chiral MEL with bulk force, the magnitude exactly matches with the external time-averaged total force. (a) Calculation procedure of a 3D magneto-dielectric core where the whole core-shell sphere is embedded in air. Core radius, $a$=600 nm, $\varepsilon_s = 8\varepsilon_0$, $\mu_s = 4\mu_0$ Bounded local immediate background (i.e. the shell) parameters: radius, $b$=800 nm and $\varepsilon_s = 4\varepsilon_0; \mu_s = 2\mu_0$. This entire core-shell is illuminated at a wavelength of $1070$ nm. (b) For plane wave illumination, $E_x = E_0 e^{i(kz-\omega t)}$: $\langle F_{out}^{Core} \rangle$ at different chirality parameters obtained from Minkowski ST at $r=a^+$ using the fields in the shell. Force $\langle F_{in}^{Core} \rangle$ based on the Chiral MEL ST at $r=a^-$ using core fields. The bulk force on the core given by other STs do not follow the trend of the total external force. (c) For circularly polarized wave illumination ($E_x + iE_y : E_x = E_0 e^{i(kz-\omega t)} = E_y$): still the bulk force by Chiral MEL ST is of the same trend as the external time-averaged total force as the chirality parameter of the core varies. (d) For linear and circularly polarized wave illumination: again our conclusions remain valid as the chirality parameter $\kappa$ of the core varies.

**S5. Why MEL and MChu formulations should better be considered as a mathematical toolkit**

Though the time-averaged modified EL and Chu formulations can resolve the 'calculation problem' [10] of time-averaged total force for real experiments, still they may better be considered just as correct mathematical toolkits and alternative approaches of the Minkowski force formulations. This is because:

(1) Such modified formulations suggest that the form of time-averaged optical bulk force inside an embedded object is dependent on the background permittivity and permeability, which is a nonlocal effect. For example- this dependency of optical bulk force inside the embedded object certainly violates the relativistic invariance [11] in instantaneous scenario. As a result, throughout the work, only time-averaged form of the modified formulations (MEL and MChu) have been discussed and their derivations in Supplement S3 and S4 have put forwarded strictly maintaining time-averaged results.

(2) Though the calculations presented in APPENDIX B in main article are time-averaged calculations, the non-mechanical term $\left\langle \frac{\partial}{\partial t} \boldsymbol{G} \right\rangle$ neither leads to the Minkowski nor the Abraham momentum of photon inside an embedded object.

(3) Let us consider that a lossless glass object is fully embedded in water. In time-averaged scenario: the modified EL force can be applied inside the embedded lossless glass scatterer to obtain the correct total time-averaged force. On the other hand, the well-known time-averaged EL force should be applied in the background water medium to determine the force felt by the background water medium. So, at two different places (i.e. at the background and inside the embedded object), we require two different forms of EL equations.

- These are the reasons why modified EL and Chu formulations should (probably) better be considered as efficient mathematical toolkits. However, Minkowski's theory of optical force is fully free from aforementioned restrictions. For example-

(a) Let us consider at first that a lossless glass object is embedded in water. Helmholtz/Minkowski's volumetric force distribution is applicable at each point of space. But in continuous medium (if lossless), it vanishes even in instantaneous scenario and appears only at the boundary of the object. But this does not mean Minkowski force is not applicable everywhere. For example- Now, we consider the glass object is absorbing. Then a local/bulk force part exists for Minkowski force along with Helmholtz's surface force. This bulk force describes how much momentum is transferring 'only' to the free carries due to local bulk fields of that glass object, which can also be obtained from internal Minkowski ST at r=a- [12].

Now, the interesting point is that: the time-averaged external Minkowski ST at r=a+ yields the time-averaged total force [Helmholtz surface force plus the bulk force on free carriers] on that embedded absorbing glass object just by employing the external fields of the scatterer (fields from water background). In addition, exactly at the object boundary, the difference of internal Minkowski ST and the external Minkowski ST leads to that surface force of Helmholtz [13,14].

When absorption takes place in background, a local/bulk force arises in Minkowski's force connected with the local fields in the background; which represents how much momentum is transferring 'only' to the free carriers of the background (but not to the embedded glass object).

- Hence Minkowski force has different physical operations in different spatial regions such as inside the absorbing embedded object; at the interface of the background and embedded object; and at the absorbing water background [13,14]. But still, it is applicable everywhere in space without any modification and also continuous for the case of instantaneous force (also relativistically form invariant [11]). Hence Minkowski ST/ force remains consistent everywhere in space for all the cases.

(b) Notably, exactly at the boundary of a generic object embedded in a material medium, the difference between the external and internal Minkowski stress tensor leads to the Helmholtz surface force [13,14]. Hence, time-averaged Minkowski ST (and Helmholtz force) should be considered valid as both external and internal ST (and force) without any modification as shown in Table 2 in the main article.